\documentclass[%
 aip,
 amsmath,amssymb,
 reprint,%
]{revtex4-2}

\usepackage{graphicx}
\usepackage{dcolumn}
\usepackage{bm}

\usepackage[utf8]{inputenc}
\usepackage[T1]{fontenc}
\usepackage{mathptmx}
\usepackage{etoolbox}

\makeatletter
\def\@email#1#2{%
 \endgroup
 \patchcmd{\titleblock@produce}
  {\frontmatter@RRAPformat}
  {\frontmatter@RRAPformat{\produce@RRAP{*#1\href{mailto:#2}{#2}}}\frontmatter@RRAPformat}
  {}{}
}%
\makeatother
\begin{document}

\preprint{AIP/123-QED}


    \renewcommand{\textfraction}{0.0}
    \renewcommand{\topfraction}{1.0}
    \renewcommand{\bottomfraction}{1.0}

    \renewcommand{\v}[1]{\bm{\mathrm{#1}}}
    \newcommand{\m}[1]{\bm{\mathsf{#1}}}
    \newcommand{\tx}[1]{\text{#1}}

\title{Uncovering the role of the density of states in controlling ultrafast spin dynamics}

\author{Martin Borchert}
\author{Clemens von Korff Schmising}%
\author{Daniel Schick}%
\author{Dieter Engel}%
\author{Sangeeta Sharma}%
\author{Sam Shallcross}
\affiliation{Max-Born-Institut f\"ur Nichtlineare Optik und Kurzzeitspektroskopie, 12489 Berlin, Germany}
\author{Stefan Eisebitt}%
\affiliation{Max-Born-Institut f\"ur Nichtlineare Optik und Kurzzeitspektroskopie, 12489 Berlin, Germany}
\affiliation{Technische Universit\"at Berlin, Institut f\"ur Optik und Atomare Physik, Strasse des 17. Juni 135, 10623 Berlin }

\date{\today}

\begin{abstract}
	At the ultrafast limit of optical spin manipulation is the theoretically predicted phenomena of optical  intersite spin transfer (OISTR), in which laser induced charge transfer between the sites of a multi-component material leads to control over magnetic order. A key prediction of this theory is that the demagnetization efficiency is determined by the availability of unoccupied states for intersite charge transfer. Employing state-of-the-art magneto-optical Kerr effect measurements with femtosecond time resolution, we probe this prediction via a systematic comparison of the ultrafast magnetic response between the 3\textit{d} ferromagnets, Fe, Co, and Ni, and their respective Pt-based alloys and multilayers. We find that (i) the demagnetization efficiency in the elemental magnets increases monotonically from Fe, via Co to Ni and, (ii), that the gain in demagnetization efficiency of the multi-component system over the pure element counterpart scales with the number of empty 3\textit{d} minority states, exactly as predicted by the OISTR effect. We support these experimental findings with \textit{ab initio} time-dependent density functional theory calculations that we show to capture the experimental trends very well.
\end{abstract}

\keywords{ultrafast magnetization dynamics}

\maketitle

Driven by the ever-increasing demands on data storage and processing, all-optical technologies are emerging as promising concepts to provide faster and more energy-efficient manipulation of units of information. The fundamental physics underpinning this technological drive is the lightwave control of magnetic order on ultrafast (femtosecond to picosecond) time scales. Research in this field has rapidly progressed since the first discovery of ultrafast demagnetization of Ni\cite{beaurepaire_ultrafast_1996}. To name only a few key discoveries, the magnetization of ferrimagnetic\cite{stanciu_all-optical_2007, radu_transient_2011} or antiferromagnetic materials\cite{olejnik_terahertz_2018} can be deterministically reversed; spin currents can be launched\cite{battiato_superdiffusive_2010, malinowski_control_2008} and employed for efficient THz emission\cite{kampfrath_terahertz_2013}; and finally the recent discovery of optically induced intersite spin transfer (OISTR)\cite{dewhurst_laser-induced_2018}.

Crucial to the technological utilization of this fundamental physics is the ability to predicatively control ultrafast spin dynamics through material properties. The OISTR effect in multi-component systems allows both ultrafast manipulation of local moments and, furthermore, posits that this control can understood and predicted from the ground state electronic spectrum, offering a possible route to designed spin dynamics properties. While the existence of the OISTR effect has now been confirmed many times in experiment (NiPt\cite{siegrist_light-wave_2019}, CoCu\cite{chen_competing_2019}, half-Heusler materials\cite{steil_efficiency_2020}, Co$_2$MnGe\cite{tengdin_direct_2020}, FeNi\cite{hofherr_ultrafast_2020}, and CoPt\cite{willems_optical_2020}), the crucial feature of the correlation of the resulting spin dynamics with features of the ground state spectrum has not, to date, been experimentally verified. A systematic exploration of a series of materials, across a large fluence range, required to establish the OISTR-predicted dependence of demagnetization on the ground state electronic spectrum thus remains an outstanding goal for an experiment.\\
In this work we aim to explore this question via a careful comparison of the demagnetization dynamics of the 3\textit{d} ferromagnets (FMs) Fe, Co, and Ni, and their respective FMPt alloys and FM$\vert$Pt multilayers (MLs) within the same experimental setup and under the same excitation conditions. According to the OISTR picture, in the FMPt alloys and FM$\vert$Pt multilayers an additional demagnetization process is available as compared to the corresponding pure elements: optical excitation of minority 5\textit{d} Pt electrons into the unoccupied minority 3\textit{d} FM channel resulting increase in the minority charge at the FM site and so, due to the lack of a compensating decrease in majority charge, a reduction of the FM site local moment. However, as the $d$-band filling increases across the transition metal series, the number of minority 3\textit{d} empty states above the Fermi energy in turn decreases, and so this process will become less effective on going from Fe via Co to Ni. We thus expect that the multi-component systems will (i) all demagnetize more efficiently than the pure 3\textit{d} elements due to the additional OISTR process and, more importantly, (ii), the demagnetization efficiency will reduce as the $d$-band fills on going from Fe via Co to Ni.

To probe these physics, we performed optical pump-probe magneto-optical Kerr effect (MOKE) measurements to determine the relative demagnetization amplitudes of all 3\textit{d} FMs and of their respective FM$\vert$Pt MLs and FMPt alloys over a wide range of excitation fluences. The ratio of the demagnetization amplitude between the pure FMs and the FM$\vert$Pt MLs and FMPt alloys was found to scale with the number of available unoccupied minority states above the Fermi level in the respective 3\textit{d} magnet, exactly as predicted by the OISTR effect. We find the largest value of the ratio for Fe to Fe$\vert$Pt, a slightly smaller one for Co to Co$\vert$Pt and a significantly smaller one for Ni to Ni$\vert$Pt MLs. We thus conclude that the experiment supports the prediction of OISTR that the early time spin dynamics in these multi-component systems is dominated by the availability of empty minority states. 
To support this conclusion we perform \emph{ab-initio} spin dynamics calculations which we show captures the experimental trends very well. We furthermore carefully probe the underlying OISTR mechanism in these materials, in particular the interplay of spin-orbit coupling and the OISTR effect.

\begin{figure}
	\centering
	\includegraphics[width=1\linewidth]{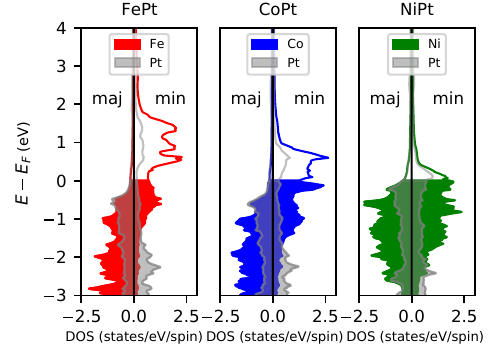}
	\caption{{Calculated density of states for 3\textit{d}-orbitals of Fe, Co, Ni, as well as the 5\textit{d}-orbitals of Pt inside the respective FMPt (50\%:50\%) ordered alloys}. During the OISTR process minority spins from Pt can transfer into the empty minority 3\textit{d}-states of the respective FM, where they compensate the majority spins and the magnetization quenches. We expect this process to be dependent on the number of available 3\textit{d}-states in the FM.}
	\label{fig:DOS}
\end{figure}

In Fig.~\ref{fig:DOS}, we display the density of states (DOS) for the three ordered alloys, FePt, CoPt, and NiPt. Shown are both occupied (indicated by the shading) and unoccupied states, as well as projections onto both the 3\textit{d} transition metal and 5\textit{d} Pt states. For each of these alloys the partially filled $d$-band results in a pronounced region of unoccupied minority states from the Fermi level ($E_\mathrm{F}$) up to $\approx2$\,eV, $\approx1$\,eV, and $\approx0.5$\,eV above $E_\mathrm{F}$ in the case of FePt, CoPt, and NiPt respectively. 
The OISTR demagnetization process -- the optical excitation of minority Pt states into minority 3\textit{d} states -- will evidently depend on the relative amounts of Pt versus FM unoccupied states: the larger the latter compared to the former the more likely is an optical transition to result in intersite charge transfer. From the ground state spectra presented in Fig.~\ref{fig:DOS}, we would thus expect OISTR to be most significant in FePt and weakest in NiPt. This behavior cannot, however, simply be deduced from the efficiency of demagnetization over a range of fluences: this will be masked by the different intrinsic magnetic properties of the transition metal elements -- the large local moment of Fe is generally more robust to demagnetization than the significantly smaller and less localized moment of Ni. To "filter" out the signatures of the OISTR pathway, we will therefore perform systematic optical pump-probe MOKE measurements comparing the magnetization dynamics of pure FM thin films and their respective FMPt alloy or FM$\vert$Pt ML thin films.

We have employed the MOKE technique in a two-color, pump-probe geometry to measure the ultrafast magnetization dynamics. Both, pump ($\lambda_\mathrm{pump} = 800$\,nm) and probe pulses ($\lambda_\mathrm{probe} = 400$\,nm) exhibit a pulse duration of 39\,fs (FWHM) leading to a measured cross-correlation of $55$\,fs, which defines the temporal resolution of the experiment. We vary the \textit{incident} fluence on the sample between 0.5\,mJ/cm$^2$ and 20\,mJ/cm$^2$ and determine the absolute values of the \textit{absorbed} fluence by careful measurements of the pump spot footprints at the sample position and by determining the transmitted (a few percent) and the reflected (50-70\%) pump power. See supplementary information for further details of the experiment.

We prepared pure FM thin films Fe(15), Co(15), and Ni(15), their respective alloys FM$_{50}$Pt$_{50}(15)$, as well as MLs Pt(2)${\vert[\mathrm{FM}(d_\mathrm{FM})\vert Pt(0.2)]\times30}$ with $d_\mathrm{Fe}=0.15$, $d_\mathrm{Co}=0.14$, and $d_\mathrm{Ni}=0.50$. All film thicknesses are given in nanometers. Note that because Ni$_{50}$Pt$_{50}$ alloys do not form a ferromagnetic phase at room temperature\cite{kawamiya_magnetic_1975}, we only investigated the FePt and CoPt alloys. All samples are magnetron-sputtered on glass wafers. The pure FMs, as well as the alloys exhibit an in-plane magnetic anisotropy and have been capped with 2\,nm of Ta to prevent oxidation. The MLs show a strong out-of-plane anisotropy and have been seeded and capped with 2\,nm of Pt.

\begin{figure*}
	\includegraphics[width=0.9\textwidth]{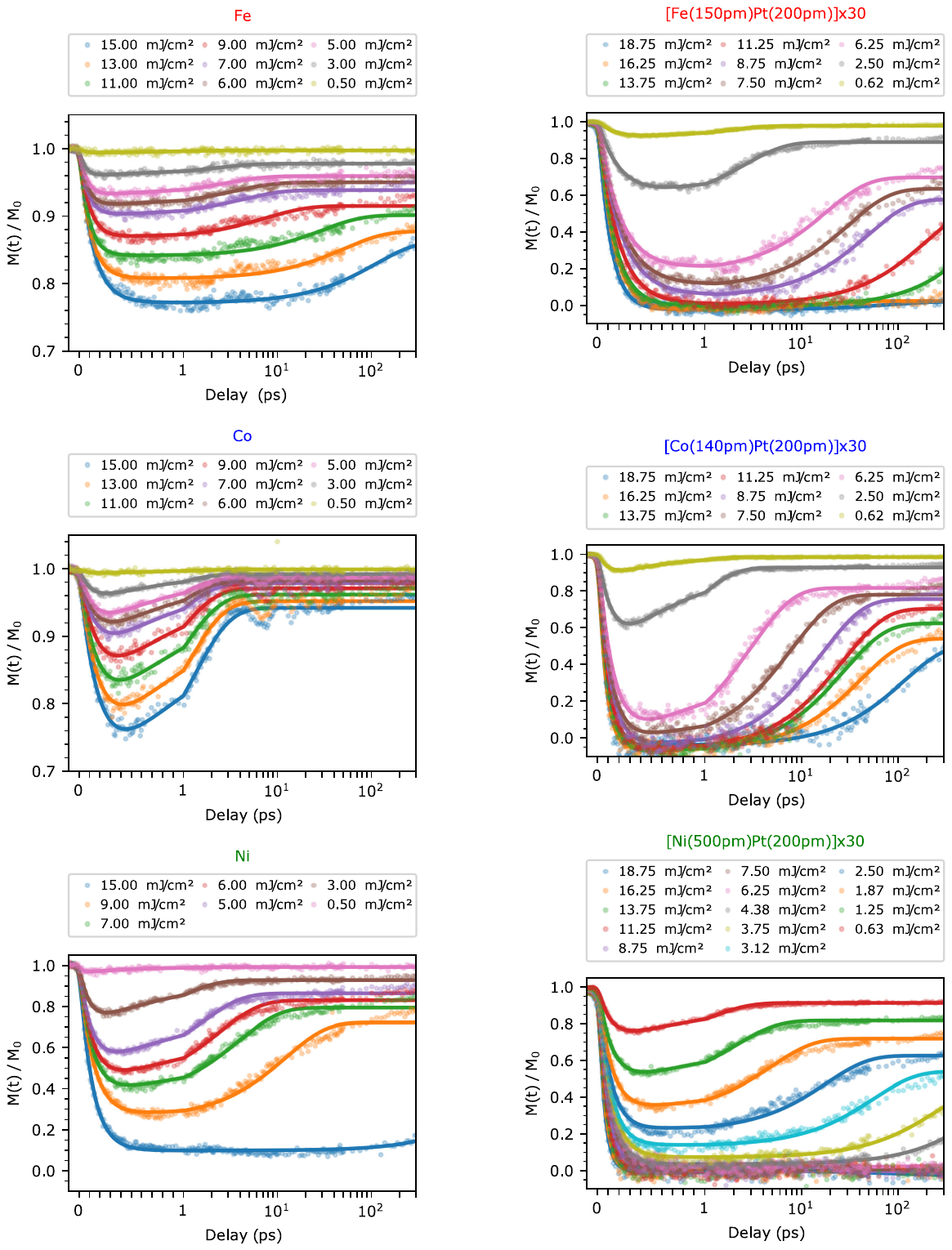}
	\caption{Normalized magnetization $M(t)/M_\mathrm{0}$ of the pure FMs and the FM$\vert$Pt multi-layers for different \textit{incident} fluences.  Note that the time axis is linear up to 1\,ps and logarithmic thereafter. Solid lines are fits with a double exponential function according to Eq.~(\ref{demageq}).}
	\label{fig:raw}
\end{figure*}

In Fig.~\ref{fig:raw}, we present the normalized demagnetization and remagnetization dynamics $M(t)/M_\mathrm{0}$ for the three pure elements and their respective multilayer systems. For the corresponding raw data of the FePt and CoPt alloys, we refer the reader to the supplementary information. 

\begin{figure}
	\centering
	\includegraphics[width=1\linewidth]{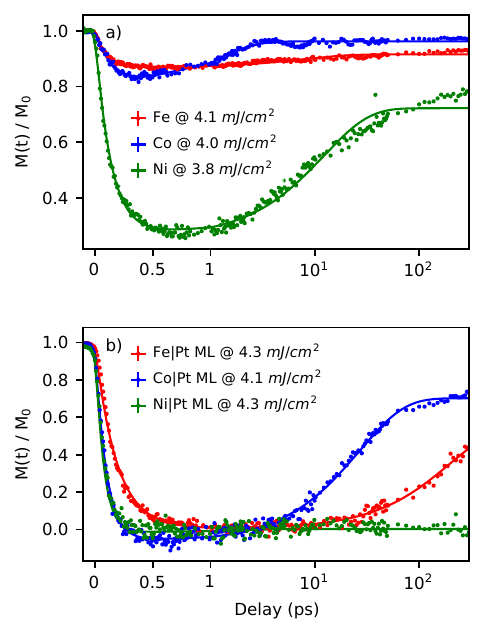}
	\caption{Normalized magnetization $M(t)/M_\mathrm{0}$ of the a) pure transition metal elements and b) FM$\vert$Pt ML for an \textit{absorbed} fluence of $\approx4$\,mJ/cm$^2$. Note that the time axis is linear up to 1\,ps and logarithmic thereafter. Solid lines are fits with a double exponential function according to Eq.~(\ref{demageq}).}
	\label{fig:absorbed_4mJ}
\end{figure}

For a rigorous investigation of the differences between these systems, we show in Fig.~\ref{fig:absorbed_4mJ} a comparison of measurements at the same \textit{absorbed} fluence of $\approx4$\,mJ/cm$^2$. Considering first the pure element thin films, one notes that despite an almost identical absorbed fluence the spin dynamics of each element is markedly different, both in terms of the maximal amplitude and temporal evolution of the magnetization. As expected, Fe presents the most robust magnetic order, demagnetizing by only approximately 10\% and exhibiting a slow recovery. Co demagnetizes somewhat more ($\approx18$\%), but rapidly recovers its magnetization (a discussion on the oscillations visible in Fig.~\ref{fig:raw} for the Co film can be found in the supplementary information). Finally Ni shows the largest response to the laser pulse, loosing almost 80\% of its initial magnetization.

In panel b) of Fig.~\ref{fig:absorbed_4mJ} the corresponding data for the FM$\vert$Pt multilayer samples are shown, again for an absorbed fluence of $\approx4$\,mJ/cm$^2$. The dramatically different spin dynamics as compared to the pure elements is immediately seen, with all multilayers now exhibiting complete demagnetization. This occurs on a different time scales for each system, with Fe fully demagnetized within $\approx 1$~ps, while Ni and Co demagnetize somewhat earlier at $\approx 0.5$\,ps. Note that the apparent demagnetization marginally in excess of 100\% for the Co$\vert$Pt multilayer likely originates in normalization errors induced by a finite DC-heating of the samples.

In order to probe the physics underpinning the spin dynamics in these systems, we require a reliable quantitative measure of the demagnetization and remagnetization amplitudes. To extract these, we fit $M(t)/M_\mathrm{0}$ by a double exponential function, convolved by a Gaussian function, $G$, with a FWHM corresponding to the measured cross-correlation of 55\,fs:

\begin{eqnarray}
	\frac{M(t)}{M_\mathrm{0}}-1 & = G * \left[H \left(A\left(e^{-t/\tau_1}-1\right)+ B\left(1-e^{-t/\tau_2}\right)\right)\right].
	\label{demageq}
\end{eqnarray}

\begin{figure}
	\centering
	\includegraphics[width=1.0\linewidth]{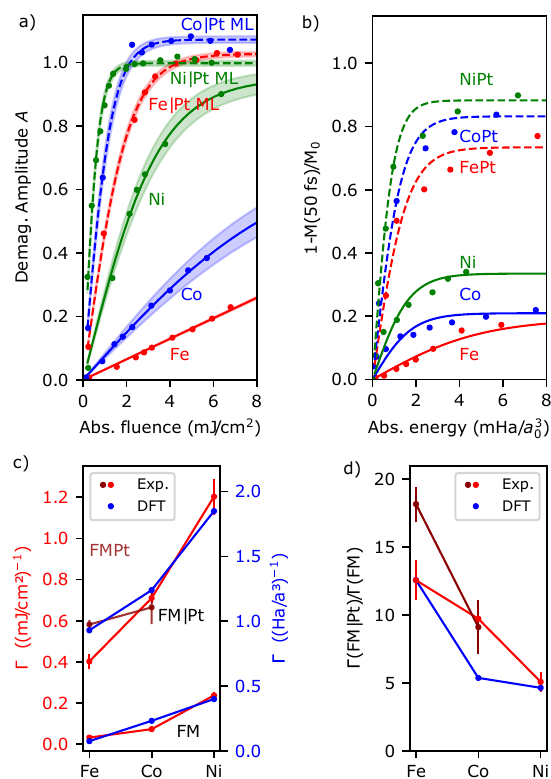}
	\caption{a) Experimental demagnetization amplitudes $A$ of the pure FMs and their respective ML as function of \textit{absorbed} fluence.  b) Results of a TDDFT calculation for the demagnetization amplitude, (1-$M$(50\,fs)$/M_\mathrm{0})$. The solid lines in both panels represent a fit of the data according to a linear-logistic function. The shaded area in a) represents one standard deviation of the fit. c) Demagnetization efficiencies, $\Gamma$, for pure FMs, FM$\vert$Pt multilayers, and FMPt alloys. d) The ratio $\Gamma$(FM$\vert$Pt)/$\Gamma$(FM) between the MLs and pure FMs exhibit a decreasing trend when going from Fe via Co to Ni,  reflecting the greater number of available states for intersite Pt-to-FM-optically-induced excitations in Fe as compared to Ni. In both panels c) and d) theoretical data from TDDFT calculations is shown alongside the experiment.}
	\label{fig:GAM}
\end{figure}

Here, $A$ and $B$ represent the demagnetization and remagnetization amplitudes, with their corresponding time constants $\tau_1$ and $\tau_2$ and $H$ is a step function centered at $t=0$. Eq.~(\ref{demageq}) provides a satisfactory description of the magnetization dynamics over the entire temporal range can be appreciated by comparison of the fitting results to the data, shown in Figs.~\ref{fig:raw} and \ref{fig:absorbed_4mJ}.
The extracted demagnetization amplitude of the multilayer and the pure elements systems as a function of the \textit{absorbed} fluence is summarized in Fig.~\ref{fig:GAM}~a): the demagnetization amplitudes of the pure elements are always lower than the corresponding multilayers, and in the case of Fe and Co substantially lower, consistent with OISTR as an additional driving mechanism for demagnetization in the multi-component systems. Importantly, the FePt and CoPt alloys exhibit an almost identical behavior to the multilayers (see supplementary information),  again in line with the OISTR mechanism, that has been shown to allow both intersite (alloy) and interface (ML) spin transfer\cite{dewhurst_substrate-controlled_2018}.

To explore the underlying role of OISTR in the spin dynamics a measure of the efficiency of the demagnetization process is required, and this is provided by the slope of the linear pre-saturation regime, see Fig.~\ref{fig:GAM}(a). We denote this slope $\Gamma$ and extract it via a fit of the demagnetization amplitude to a linear function that transitions to a logistic function above its inflection point. This is presented in Fig.~\ref{fig:GAM}~c) and summarized in Tab.~\ref{tab:demagnetisation_efficiency}, where one can see that demagnetization efficiency (i) increases on going from Fe via Co to Ni and (ii) is substantially larger for the alloys and ML, compared to the pure elements.

At this point it is important to draw attention to the observation that the trend of demagnetization efficiency of the pure element systems reported here evidently does not follow the Curie temperatures, $T_\mathrm{c}$, as generally assumed by the phenomenological three temperature model, but rather the magnitude of the ground state moment\cite{radu_ultrafast_2015}. Calculations based on the three temperature model\cite{koopmans_explaining_2010}, using tabulated values of the material constants, predict a more efficient demagnetization of Fe compared to Co, which we do not observe.
Thus, we will restrict further analysis of multi-component versus pure element systems to the OISTR mechanism.
We further note that, to our own surprise and to the best of our knowledge, this is the first time that all three 3\textit{d} transition metals, Fe, Co, and Ni are compared in a quantitative and systematic experimental study under the same excitation conditions.

\begin{table}[h]
	\begin{tabular}{|c|c|}
		\hline
		& $\Gamma$                         \\ 
		& \%/(mJ/cm$^2$)                    \\ \hline
		Fe          & 3.2 $\pm$ 0.1         \\ \hline
		Co          & 7.3 $\pm$ 0.7         \\ \hline
		Ni          & 23.7 $\pm$ 1.7        \\ \hline
		FePt        & 58.1$\pm$ 2.3         \\ \hline
		CoPt        & 66.5 $\pm$ 8.2           \\ \hline
		Fe$\vert$Pt ML & 40.2 $\pm$ 3.5     \\ \hline
		Co$\vert$Pt ML & 71.0 $\pm$ 2.3     \\ \hline
		Ni$\vert$Pt ML & 120.3 $\pm$ 8.7     \\ \hline
	\end{tabular}
	\caption{Experimental values of the demagnetization efficiency, $\Gamma$, per \textit{absorbed} fluence in mJ/cm$^2$. Values and standard deviations are extracted from the linear-logistic fits as described in the main text.}
	\label{tab:demagnetisation_efficiency}
\end{table}

To discern the underlying OISTR physics present in the multilayer and alloy samples we must isolate the intrinsic contribution to spin dynamics from the transition metal. To this end, we form the ratio of the demagnetization efficiencies for the pure transition metal element and the corresponding multi-component system: $\Gamma(\mathrm{FM\vert Pt})/\Gamma(\mathrm{FM})$. This ratio, that encodes the \emph{change} in demagnetization efficiency on going from the pure element to the multilayer or alloys, is shown in Fig.~\ref{fig:GAM}~d). One notices that Fe shows the greatest increase in demagnetization efficiency and Ni the least, exactly as expected from the OISTR mechanism. Finally, comparison of the demagnetization constant, $\tau_1$, reveals an accelerated rate for the multi-component systems compared to the pure elements, which further corroborates the important role of OISTR (see supplementary information).

Our results thus point towards two conclusions: firstly, that the FM$\vert$Pt multilayers, as well as the FMPt alloys, demagnetize significantly more efficiently than the corresponding pure FMs; secondly, that the trend of increase of the demagnetization efficiency upon forming Pt based alloys and multilayers from the pure elements scales with the available states above the FERMI energy and is therefore consistent with the OISTR mechanism.

In order to further substantiate our experimental findings, we now perform time dependent density function theory (TDDFT) calculations of: (i) bulk elements and (ii) ordered L1$_0$ alloys. Note that while the experiment considers either solid solution alloys or thin film multilayers, the similar demagnetization amplitude found for these systems justifies the use of ordered alloys to represent FM$\vert$Pt multi-component systems in theory. 

We consider a laser pump pulse with central frequency 1.55\,eV, a pulse duration of 12\,fs (FWHM) and an incident fluence of 36\,mJ/cm$^2$. All calculations were performed within the framework of the all-electron full-potential linearized augmented plane wave method\cite{singh}, as implemented in the Elk code\cite{elk,dewhurst2016} and used the adiabatic local density approximation (see Ref.~\onlinecite{dewhurst2016} and Supplementary Information for full details of the method). To access the demagnetization amplitude we consider the ratio between the magnetic moment after the laser pulse and the ground state magnetic moment: $1-M(50~\text{fs})/M_\mathrm{0}$. This is shown in Fig.~\ref{fig:GAM}~b), where it can be seen that the demagnetization amplitude calculated in this way very closely reproduces the experiment trends, not only for the pure elements but also for the multi-component systems. Extracting the gradient of the linear dependence on fluence, $\Gamma$, exactly as in experiment by fitting to a linear logistic function, we find the demagnetization efficiencies shown in Fig.~\ref{fig:GAM}~c). Evidently these capture the trends of the demagnetization efficiencies very well, again for both the pure elements and multi-component systems. Taking the ratio to access the gain in demagnetization efficiency on going from the pure element to the multi-component system, we find remarkably good agreement with the trend seen in experiment, see Fig.~\ref{fig:GAM}~d). The lower saturation demagnetization as compared to experiment reflects the restriction to electronic excitations in the simulation -- optical excitations and spin-orbit induced transitions between majority and minority -- while in experiment the longer time scales involved certainly imply the involvement of the lattice in the spin dynamics.
The monotonically decreasing trend from Fe via Co to Ni matches precisely with the expectation of the underlying OISTR effect -- namely that as the $d$-band fills on going from Fe via Co to Ni less unoccupied states are available for Pt to FM intersite spin transfer. 

\begin{figure}
	\centering
	\includegraphics[width=1.0\linewidth]{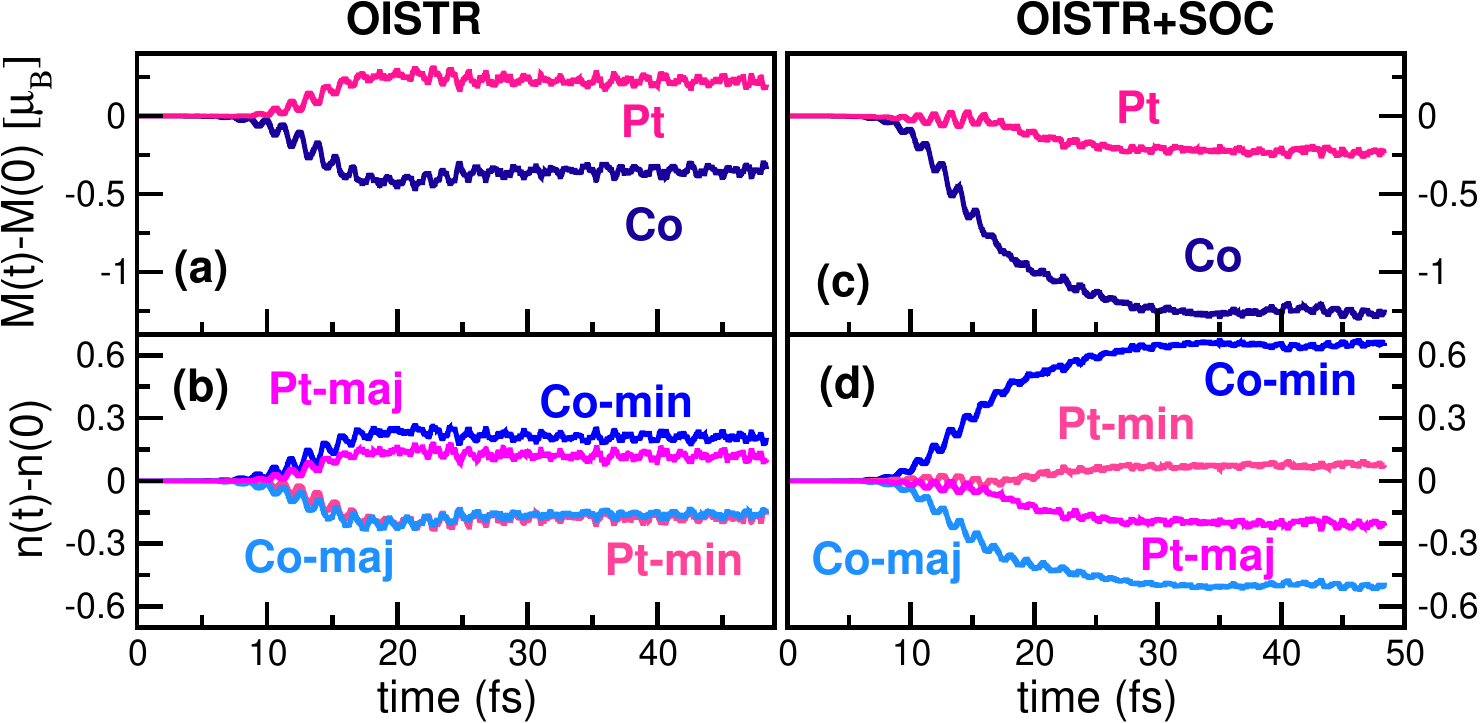}
	\caption{Transient atom-resolved magnetic moment for laser pumped CoPt (a) without SOC and (c) in presence of SOC term in the Hamiltonian. Transient atom resolved majority and minority charge in laser pumped CoPt (b) without SOC and (d) in presence of SOC term in the Hamiltonian. }
	\label{fig:t1}
\end{figure}

In the following, we examine the CoPt alloy in more detail, and further probe the microscopic mechanism involved. By switching off spin-orbit coupling we retain only the optically driven OISTR effect; the resulting dependence of the moment and the Pt and Co majority and minority charge as shown in Fig.~\ref{fig:t1}~a,b). The loss in Co moment is almost perfectly compensated by the gain in Pt -- exactly as expected for an optically induced transfer of charge between Pt and Co. This situation changes on switching on spin-orbit coupling. There are now two \emph{co-occurring} processes: SO induced transitions between majority and minority on Pt and OISTR, and thus a pathway for \emph{both} Pt majority and Pt minority to Co minority resulting in increased demagnetization, as can be seen in Fig.~\ref{fig:t1}~c,d).\\
Finally, we briefly discuss the longer time scales considered in the experiment compared to the theory. The calculations describe the very early time regime in which electronic effects dominate the dynamics, namely optical and spin-orbit induced transitions. Importantly, in previous experiments \emph{also} within this very early time regime\cite{siegrist_light-wave_2019} the agreement with TDDFT has been shown to be remarkably quantitative, demonstrating that this methodology fully captures femtoscale spin dynamics. However, for longer times scales several physical effects, which are not included in the TDDFT calculations, can become important: (i) coupling of the electronic degrees of freedom to the nuclear degrees of freedom and radiative electromagnetic coupling, (ii) large length scale magnetization non-colinearity (e.g. magnons) or large length scale currents (such as described by the super-diffusive mechanism). Nevertheless, the experimental observation that a decreasing trend of $\Gamma$(FMPt)/$\Gamma$(FM) when going from Fe via Co to Ni coincides with the decreasing number of available states presents very compelling evidence of OISTR.  The good agreement of the experimental results with early time regime TDDFT calculations then also suggests that OISTR contributes substantially to the demagnetization, dictating the response even after hundreds of femtoseconds.\\
In this work we have addressed the question of whether the demagnetization dynamics of multilayers and alloys of the 3\textit{d} ferromagnets (FM) Fe, Co, and Ni with Pt is governed by the availability of states for laser induced transitions from occupied Pt to unoccupied minority FM states, as expected on the basis of the OISTR effect. By carefully determining the gain in demagnetization efficiency on going from the pure element to the multi-component system, we show that the resulting trend of a decrease in efficiency gain from Fe via Co to Ni is completely consistent with the OISTR effect.
While previous work has shown that the OISTR effect is responsible for increased moment loss in multi-component systems as compared to their pure element counterparts
\cite{siegrist_light-wave_2019,chen_competing_2019,steil_efficiency_2020,
tengdin_direct_2020,hofherr_ultrafast_2020,willems_optical_2020}, our work goes further in showing that the magnitude of this effect, probed over a range of materials, follows the behavior expected from an availability-of-states argument based on the ground state spectrum.
Our experimental findings thus highlight the quantitative and predictive aspect of the dependency of OISTR-governed spin dynamics on the number of available unoccupied states, potentially opening up pathways for tailoring demagnetization dynamics based on density of states engineering.

\section*{Acknowledgements}
This project was supported by the Deutsche Forschungsgemeinschaft through the Collaborative Research Center TRR~227 on Ultrafast Spin Dynamics, project A02 and A04. S.~Shallcross would like to thank DFG for funding through SH498/4-1. The authors acknowledge the North-German Supercomputing Alliance (HLRN) for providing HPC resources that have contributed to the research results reported in this paper.

\section*{Conflict of Interest}
The authors report no conflicts of interest.
\section*{Data Availability}
The data that support the findings of this study are available from the corresponding author upon reasonable request

\section*{References}
\bibliography{z_bib}

\begin{thebibliography}{29}%
\makeatletter
\providecommand \@ifxundefined [1]{%
 \@ifx{#1\undefined}
}%
\providecommand \@ifnum [1]{%
 \ifnum #1\expandafter \@firstoftwo
 \else \expandafter \@secondoftwo
 \fi
}%
\providecommand \@ifx [1]{%
 \ifx #1\expandafter \@firstoftwo
 \else \expandafter \@secondoftwo
 \fi
}%
\providecommand \natexlab [1]{#1}%
\providecommand \enquote  [1]{``#1''}%
\providecommand \bibnamefont  [1]{#1}%
\providecommand \bibfnamefont [1]{#1}%
\providecommand \citenamefont [1]{#1}%
\providecommand \href@noop [0]{\@secondoftwo}%
\providecommand \href [0]{\begingroup \@sanitize@url \@href}%
\providecommand \@href[1]{\@@startlink{#1}\@@href}%
\providecommand \@@href[1]{\endgroup#1\@@endlink}%
\providecommand \@sanitize@url [0]{\catcode `\\12\catcode `\$12\catcode
  `\&12\catcode `\#12\catcode `\^12\catcode `\_12\catcode `\%12\relax}%
\providecommand \@@startlink[1]{}%
\providecommand \@@endlink[0]{}%
\providecommand \url  [0]{\begingroup\@sanitize@url \@url }%
\providecommand \@url [1]{\endgroup\@href {#1}{\urlprefix }}%
\providecommand \urlprefix  [0]{URL }%
\providecommand \Eprint [0]{\href }%
\providecommand \doibase [0]{https://doi.org/}%
\providecommand \selectlanguage [0]{\@gobble}%
\providecommand \bibinfo  [0]{\@secondoftwo}%
\providecommand \bibfield  [0]{\@secondoftwo}%
\providecommand \translation [1]{[#1]}%
\providecommand \BibitemOpen [0]{}%
\providecommand \bibitemStop [0]{}%
\providecommand \bibitemNoStop [0]{.\EOS\space}%
\providecommand \EOS [0]{\spacefactor3000\relax}%
\providecommand \BibitemShut  [1]{\csname bibitem#1\endcsname}%
\let\auto@bib@innerbib\@empty
\bibitem [{\citenamefont {Beaurepaire}\ \emph {et~al.}(1996)\citenamefont
  {Beaurepaire}, \citenamefont {Merle}, \citenamefont {Daunois},\ and\
  \citenamefont {Bigot}}]{beaurepaire_ultrafast_1996}%
  \BibitemOpen
  \bibfield  {author} {\bibinfo {author} {\bibfnamefont {E.}~\bibnamefont
  {Beaurepaire}}, \bibinfo {author} {\bibfnamefont {J.-C.}\ \bibnamefont
  {Merle}}, \bibinfo {author} {\bibfnamefont {A.}~\bibnamefont {Daunois}},\
  and\ \bibinfo {author} {\bibfnamefont {J.-Y.}\ \bibnamefont {Bigot}},\
  }\bibfield  {title} {\enquote {\bibinfo {title} {Ultrafast {Spin} {Dynamics}
  in {Ferromagnetic} {Nickel}},}\ }\href
  {https://doi.org/10.1103/PhysRevLett.76.4250} {\bibfield  {journal} {\bibinfo
   {journal} {Phys. Rev. Lett.}\ }\textbf {\bibinfo {volume} {76}},\ \bibinfo
  {pages} {4250--4253} (\bibinfo {year} {1996})}\BibitemShut {NoStop}%
\bibitem [{\citenamefont {Stanciu}\ \emph {et~al.}(2007)\citenamefont
  {Stanciu}, \citenamefont {Hansteen}, \citenamefont {Kimel}, \citenamefont
  {Kirilyuk}, \citenamefont {Tsukamoto}, \citenamefont {Itoh},\ and\
  \citenamefont {Rasing}}]{stanciu_all-optical_2007}%
  \BibitemOpen
  \bibfield  {author} {\bibinfo {author} {\bibfnamefont {C.~D.}\ \bibnamefont
  {Stanciu}}, \bibinfo {author} {\bibfnamefont {F.}~\bibnamefont {Hansteen}},
  \bibinfo {author} {\bibfnamefont {A.~V.}\ \bibnamefont {Kimel}}, \bibinfo
  {author} {\bibfnamefont {A.}~\bibnamefont {Kirilyuk}}, \bibinfo {author}
  {\bibfnamefont {A.}~\bibnamefont {Tsukamoto}}, \bibinfo {author}
  {\bibfnamefont {A.}~\bibnamefont {Itoh}},\ and\ \bibinfo {author}
  {\bibfnamefont {T.}~\bibnamefont {Rasing}},\ }\bibfield  {title} {\enquote
  {\bibinfo {title} {All-{Optical} {Magnetic} {Recording} with {Circularly}
  {Polarized} {Light}},}\ }\href
  {https://doi.org/10.1103/PhysRevLett.99.047601} {\bibfield  {journal}
  {\bibinfo  {journal} {Phys. Rev. Lett.}\ }\textbf {\bibinfo {volume} {99}},\
  \bibinfo {pages} {047601} (\bibinfo {year} {2007})}\BibitemShut {NoStop}%
\bibitem [{\citenamefont {Radu}\ \emph {et~al.}(2011)\citenamefont {Radu},
  \citenamefont {Vahaplar}, \citenamefont {Stamm}, \citenamefont {Kachel},
  \citenamefont {Pontius}, \citenamefont {Dürr}, \citenamefont {Ostler},
  \citenamefont {Barker}, \citenamefont {Evans}, \citenamefont {Chantrell},
  \citenamefont {Tsukamoto}, \citenamefont {Itoh}, \citenamefont {Kirilyuk},
  \citenamefont {Rasing},\ and\ \citenamefont {Kimel}}]{radu_transient_2011}%
  \BibitemOpen
  \bibfield  {author} {\bibinfo {author} {\bibfnamefont {I.}~\bibnamefont
  {Radu}}, \bibinfo {author} {\bibfnamefont {K.}~\bibnamefont {Vahaplar}},
  \bibinfo {author} {\bibfnamefont {C.}~\bibnamefont {Stamm}}, \bibinfo
  {author} {\bibfnamefont {T.}~\bibnamefont {Kachel}}, \bibinfo {author}
  {\bibfnamefont {N.}~\bibnamefont {Pontius}}, \bibinfo {author} {\bibfnamefont
  {H.~A.}\ \bibnamefont {Dürr}}, \bibinfo {author} {\bibfnamefont {T.~A.}\
  \bibnamefont {Ostler}}, \bibinfo {author} {\bibfnamefont {J.}~\bibnamefont
  {Barker}}, \bibinfo {author} {\bibfnamefont {R.~F.~L.}\ \bibnamefont
  {Evans}}, \bibinfo {author} {\bibfnamefont {R.~W.}\ \bibnamefont
  {Chantrell}}, \bibinfo {author} {\bibfnamefont {A.}~\bibnamefont
  {Tsukamoto}}, \bibinfo {author} {\bibfnamefont {A.}~\bibnamefont {Itoh}},
  \bibinfo {author} {\bibfnamefont {A.}~\bibnamefont {Kirilyuk}}, \bibinfo
  {author} {\bibfnamefont {T.}~\bibnamefont {Rasing}},\ and\ \bibinfo {author}
  {\bibfnamefont {A.~V.}\ \bibnamefont {Kimel}},\ }\bibfield  {title} {\enquote
  {\bibinfo {title} {Transient ferromagnetic-like state mediating ultrafast
  reversal of antiferromagnetically coupled spins},}\ }\href
  {https://doi.org/10.1038/nature09901} {\bibfield  {journal} {\bibinfo
  {journal} {Nature}\ }\textbf {\bibinfo {volume} {472}},\ \bibinfo {pages}
  {205--208} (\bibinfo {year} {2011})}\BibitemShut {NoStop}%
\bibitem [{\citenamefont {Olejník}\ \emph {et~al.}(2018)\citenamefont
  {Olejník}, \citenamefont {Seifert}, \citenamefont {Kašpar}, \citenamefont
  {Novák}, \citenamefont {Wadley}, \citenamefont {Campion}, \citenamefont
  {Baumgartner}, \citenamefont {Gambardella}, \citenamefont {Němec},
  \citenamefont {Wunderlich}, \citenamefont {Sinova}, \citenamefont {Kužel},
  \citenamefont {Müller}, \citenamefont {Kampfrath},\ and\ \citenamefont
  {Jungwirth}}]{olejnik_terahertz_2018}%
  \BibitemOpen
  \bibfield  {author} {\bibinfo {author} {\bibfnamefont {K.}~\bibnamefont
  {Olejník}}, \bibinfo {author} {\bibfnamefont {T.}~\bibnamefont {Seifert}},
  \bibinfo {author} {\bibfnamefont {Z.}~\bibnamefont {Kašpar}}, \bibinfo
  {author} {\bibfnamefont {V.}~\bibnamefont {Novák}}, \bibinfo {author}
  {\bibfnamefont {P.}~\bibnamefont {Wadley}}, \bibinfo {author} {\bibfnamefont
  {R.~P.}\ \bibnamefont {Campion}}, \bibinfo {author} {\bibfnamefont
  {M.}~\bibnamefont {Baumgartner}}, \bibinfo {author} {\bibfnamefont
  {P.}~\bibnamefont {Gambardella}}, \bibinfo {author} {\bibfnamefont
  {P.}~\bibnamefont {Němec}}, \bibinfo {author} {\bibfnamefont
  {J.}~\bibnamefont {Wunderlich}}, \bibinfo {author} {\bibfnamefont
  {J.}~\bibnamefont {Sinova}}, \bibinfo {author} {\bibfnamefont
  {P.}~\bibnamefont {Kužel}}, \bibinfo {author} {\bibfnamefont
  {M.}~\bibnamefont {Müller}}, \bibinfo {author} {\bibfnamefont
  {T.}~\bibnamefont {Kampfrath}},\ and\ \bibinfo {author} {\bibfnamefont
  {T.}~\bibnamefont {Jungwirth}},\ }\bibfield  {title} {\enquote {\bibinfo
  {title} {Terahertz electrical writing speed in an antiferromagnetic
  memory},}\ }\href {https://doi.org/10.1126/sciadv.aar3566} {\bibfield
  {journal} {\bibinfo  {journal} {Sci. Adv.}\ }\textbf {\bibinfo {volume}
  {4}},\ \bibinfo {pages} {eaar3566} (\bibinfo {year} {2018})}\BibitemShut
  {NoStop}%
\bibitem [{\citenamefont {Battiato}, \citenamefont {Carva},\ and\ \citenamefont
  {Oppeneer}(2010)}]{battiato_superdiffusive_2010}%
  \BibitemOpen
  \bibfield  {author} {\bibinfo {author} {\bibfnamefont {M.}~\bibnamefont
  {Battiato}}, \bibinfo {author} {\bibfnamefont {K.}~\bibnamefont {Carva}},\
  and\ \bibinfo {author} {\bibfnamefont {P.~M.}\ \bibnamefont {Oppeneer}},\
  }\bibfield  {title} {\enquote {\bibinfo {title} {Superdiffusive {Spin}
  {Transport} as a {Mechanism} of {Ultrafast} {Demagnetization}},}\ }\href
  {https://doi.org/10.1103/PhysRevLett.105.027203} {\bibfield  {journal}
  {\bibinfo  {journal} {Phys. Rev. Lett.}\ }\textbf {\bibinfo {volume} {105}},\
  \bibinfo {pages} {027203} (\bibinfo {year} {2010})}\BibitemShut {NoStop}%
\bibitem [{\citenamefont {Malinowski}\ \emph {et~al.}(2008)\citenamefont
  {Malinowski}, \citenamefont {Dalla~Longa}, \citenamefont {Rietjens},
  \citenamefont {Paluskar}, \citenamefont {Huijink}, \citenamefont {Swagten},\
  and\ \citenamefont {Koopmans}}]{malinowski_control_2008}%
  \BibitemOpen
  \bibfield  {author} {\bibinfo {author} {\bibfnamefont {G.}~\bibnamefont
  {Malinowski}}, \bibinfo {author} {\bibfnamefont {F.}~\bibnamefont
  {Dalla~Longa}}, \bibinfo {author} {\bibfnamefont {J.~H.~H.}\ \bibnamefont
  {Rietjens}}, \bibinfo {author} {\bibfnamefont {P.~V.}\ \bibnamefont
  {Paluskar}}, \bibinfo {author} {\bibfnamefont {R.}~\bibnamefont {Huijink}},
  \bibinfo {author} {\bibfnamefont {H.~J.~M.}\ \bibnamefont {Swagten}},\ and\
  \bibinfo {author} {\bibfnamefont {B.}~\bibnamefont {Koopmans}},\ }\bibfield
  {title} {\enquote {\bibinfo {title} {Control of speed and efficiency of
  ultrafast demagnetization by direct transfer of spin angular momentum},}\
  }\href {https://doi.org/10.1038/nphys1092} {\bibfield  {journal} {\bibinfo
  {journal} {Nature Physics}\ }\textbf {\bibinfo {volume} {4}},\ \bibinfo
  {pages} {855--858} (\bibinfo {year} {2008})}\BibitemShut {NoStop}%
\bibitem [{\citenamefont {Kampfrath}\ \emph {et~al.}(2013)\citenamefont
  {Kampfrath}, \citenamefont {Battiato}, \citenamefont {Maldonado},
  \citenamefont {Eilers}, \citenamefont {Nötzold}, \citenamefont {Mährlein},
  \citenamefont {Zbarsky}, \citenamefont {Freimuth}, \citenamefont {Mokrousov},
  \citenamefont {Blügel}, \citenamefont {Wolf}, \citenamefont {Radu},
  \citenamefont {Oppeneer},\ and\ \citenamefont
  {Münzenberg}}]{kampfrath_terahertz_2013}%
  \BibitemOpen
  \bibfield  {author} {\bibinfo {author} {\bibfnamefont {T.}~\bibnamefont
  {Kampfrath}}, \bibinfo {author} {\bibfnamefont {M.}~\bibnamefont {Battiato}},
  \bibinfo {author} {\bibfnamefont {P.}~\bibnamefont {Maldonado}}, \bibinfo
  {author} {\bibfnamefont {G.}~\bibnamefont {Eilers}}, \bibinfo {author}
  {\bibfnamefont {J.}~\bibnamefont {Nötzold}}, \bibinfo {author}
  {\bibfnamefont {S.}~\bibnamefont {Mährlein}}, \bibinfo {author}
  {\bibfnamefont {V.}~\bibnamefont {Zbarsky}}, \bibinfo {author} {\bibfnamefont
  {F.}~\bibnamefont {Freimuth}}, \bibinfo {author} {\bibfnamefont
  {Y.}~\bibnamefont {Mokrousov}}, \bibinfo {author} {\bibfnamefont
  {S.}~\bibnamefont {Blügel}}, \bibinfo {author} {\bibfnamefont
  {M.}~\bibnamefont {Wolf}}, \bibinfo {author} {\bibfnamefont {I.}~\bibnamefont
  {Radu}}, \bibinfo {author} {\bibfnamefont {P.~M.}\ \bibnamefont {Oppeneer}},\
  and\ \bibinfo {author} {\bibfnamefont {M.}~\bibnamefont {Münzenberg}},\
  }\bibfield  {title} {\enquote {\bibinfo {title} {Terahertz spin current
  pulses controlled by magnetic heterostructures},}\ }\href
  {https://doi.org/10.1038/nnano.2013.43} {\bibfield  {journal} {\bibinfo
  {journal} {Nature Nanotechnology}\ }\textbf {\bibinfo {volume} {8}},\
  \bibinfo {pages} {256--260} (\bibinfo {year} {2013})}\BibitemShut {NoStop}%
\bibitem [{\citenamefont {Dewhurst}\ \emph
  {et~al.}(2018{\natexlab{a}})\citenamefont {Dewhurst}, \citenamefont
  {Elliott}, \citenamefont {Shallcross}, \citenamefont {Gross},\ and\
  \citenamefont {Sharma}}]{dewhurst_laser-induced_2018}%
  \BibitemOpen
  \bibfield  {author} {\bibinfo {author} {\bibfnamefont {J.~K.}\ \bibnamefont
  {Dewhurst}}, \bibinfo {author} {\bibfnamefont {P.}~\bibnamefont {Elliott}},
  \bibinfo {author} {\bibfnamefont {S.}~\bibnamefont {Shallcross}}, \bibinfo
  {author} {\bibfnamefont {E.~K.~U.}\ \bibnamefont {Gross}},\ and\ \bibinfo
  {author} {\bibfnamefont {S.}~\bibnamefont {Sharma}},\ }\bibfield  {title}
  {\enquote {\bibinfo {title} {Laser-{Induced} {Intersite} {Spin}
  {Transfer}},}\ }\href {https://doi.org/10.1021/acs.nanolett.7b05118}
  {\bibfield  {journal} {\bibinfo  {journal} {Nano Lett.}\ }\textbf {\bibinfo
  {volume} {18}},\ \bibinfo {pages} {1842--1848} (\bibinfo {year}
  {2018}{\natexlab{a}})}\BibitemShut {NoStop}%
\bibitem [{\citenamefont {Siegrist}\ \emph {et~al.}(2019)\citenamefont
  {Siegrist}, \citenamefont {Gessner}, \citenamefont {Ossiander}, \citenamefont
  {Denker}, \citenamefont {Chang}, \citenamefont {Schröder}, \citenamefont
  {Guggenmos}, \citenamefont {Cui}, \citenamefont {Walowski}, \citenamefont
  {Martens}, \citenamefont {Dewhurst}, \citenamefont {Kleineberg},
  \citenamefont {Münzenberg}, \citenamefont {Sharma},\ and\ \citenamefont
  {Schultze}}]{siegrist_light-wave_2019}%
  \BibitemOpen
  \bibfield  {author} {\bibinfo {author} {\bibfnamefont {F.}~\bibnamefont
  {Siegrist}}, \bibinfo {author} {\bibfnamefont {J.~A.}\ \bibnamefont
  {Gessner}}, \bibinfo {author} {\bibfnamefont {M.}~\bibnamefont {Ossiander}},
  \bibinfo {author} {\bibfnamefont {C.}~\bibnamefont {Denker}}, \bibinfo
  {author} {\bibfnamefont {Y.-P.}\ \bibnamefont {Chang}}, \bibinfo {author}
  {\bibfnamefont {M.~C.}\ \bibnamefont {Schröder}}, \bibinfo {author}
  {\bibfnamefont {A.}~\bibnamefont {Guggenmos}}, \bibinfo {author}
  {\bibfnamefont {Y.}~\bibnamefont {Cui}}, \bibinfo {author} {\bibfnamefont
  {J.}~\bibnamefont {Walowski}}, \bibinfo {author} {\bibfnamefont
  {U.}~\bibnamefont {Martens}}, \bibinfo {author} {\bibfnamefont {J.~K.}\
  \bibnamefont {Dewhurst}}, \bibinfo {author} {\bibfnamefont {U.}~\bibnamefont
  {Kleineberg}}, \bibinfo {author} {\bibfnamefont {M.}~\bibnamefont
  {Münzenberg}}, \bibinfo {author} {\bibfnamefont {S.}~\bibnamefont
  {Sharma}},\ and\ \bibinfo {author} {\bibfnamefont {M.}~\bibnamefont
  {Schultze}},\ }\bibfield  {title} {\enquote {\bibinfo {title} {Light-wave
  dynamic control of magnetism},}\ }\href
  {https://doi.org/10.1038/s41586-019-1333-x} {\bibfield  {journal} {\bibinfo
  {journal} {Nature}\ }\textbf {\bibinfo {volume} {571}},\ \bibinfo {pages}
  {240--244} (\bibinfo {year} {2019})}\BibitemShut {NoStop}%
\bibitem [{\citenamefont {Chen}\ \emph {et~al.}(2019)\citenamefont {Chen},
  \citenamefont {Bovensiepen}, \citenamefont {Eschenlohr}, \citenamefont
  {Müller}, \citenamefont {Elliott}, \citenamefont {Gross}, \citenamefont
  {Dewhurst},\ and\ \citenamefont {Sharma}}]{chen_competing_2019}%
  \BibitemOpen
  \bibfield  {author} {\bibinfo {author} {\bibfnamefont {J.}~\bibnamefont
  {Chen}}, \bibinfo {author} {\bibfnamefont {U.}~\bibnamefont {Bovensiepen}},
  \bibinfo {author} {\bibfnamefont {A.}~\bibnamefont {Eschenlohr}}, \bibinfo
  {author} {\bibfnamefont {T.}~\bibnamefont {Müller}}, \bibinfo {author}
  {\bibfnamefont {P.}~\bibnamefont {Elliott}}, \bibinfo {author} {\bibfnamefont
  {E.}~\bibnamefont {Gross}}, \bibinfo {author} {\bibfnamefont
  {J.}~\bibnamefont {Dewhurst}},\ and\ \bibinfo {author} {\bibfnamefont
  {S.}~\bibnamefont {Sharma}},\ }\bibfield  {title} {\enquote {\bibinfo {title}
  {Competing {Spin} {Transfer} and {Dissipation} at {Co}/{Cu}(001) {Interfaces}
  on {Femtosecond} {Timescales}},}\ }\href
  {https://doi.org/10.1103/PhysRevLett.122.067202} {\bibfield  {journal}
  {\bibinfo  {journal} {Phys. Rev. Lett.}\ }\textbf {\bibinfo {volume} {122}},\
  \bibinfo {pages} {067202} (\bibinfo {year} {2019})}\BibitemShut {NoStop}%
\bibitem [{\citenamefont {Steil}\ \emph {et~al.}(2020)\citenamefont {Steil},
  \citenamefont {Walowski}, \citenamefont {Gerhard}, \citenamefont {Kiessling},
  \citenamefont {Ebke}, \citenamefont {Thomas}, \citenamefont {Kubota},
  \citenamefont {Oogane}, \citenamefont {Ando}, \citenamefont {Otto},
  \citenamefont {Mann}, \citenamefont {Hofherr}, \citenamefont {Elliott},
  \citenamefont {Dewhurst}, \citenamefont {Reiss}, \citenamefont {Molenkamp},
  \citenamefont {Aeschlimann}, \citenamefont {Cinchetti}, \citenamefont
  {Münzenberg}, \citenamefont {Sharma},\ and\ \citenamefont
  {Mathias}}]{steil_efficiency_2020}%
  \BibitemOpen
  \bibfield  {author} {\bibinfo {author} {\bibfnamefont {D.}~\bibnamefont
  {Steil}}, \bibinfo {author} {\bibfnamefont {J.}~\bibnamefont {Walowski}},
  \bibinfo {author} {\bibfnamefont {F.}~\bibnamefont {Gerhard}}, \bibinfo
  {author} {\bibfnamefont {T.}~\bibnamefont {Kiessling}}, \bibinfo {author}
  {\bibfnamefont {D.}~\bibnamefont {Ebke}}, \bibinfo {author} {\bibfnamefont
  {A.}~\bibnamefont {Thomas}}, \bibinfo {author} {\bibfnamefont
  {T.}~\bibnamefont {Kubota}}, \bibinfo {author} {\bibfnamefont
  {M.}~\bibnamefont {Oogane}}, \bibinfo {author} {\bibfnamefont
  {Y.}~\bibnamefont {Ando}}, \bibinfo {author} {\bibfnamefont {J.}~\bibnamefont
  {Otto}}, \bibinfo {author} {\bibfnamefont {A.}~\bibnamefont {Mann}}, \bibinfo
  {author} {\bibfnamefont {M.}~\bibnamefont {Hofherr}}, \bibinfo {author}
  {\bibfnamefont {P.}~\bibnamefont {Elliott}}, \bibinfo {author} {\bibfnamefont
  {J.~K.}\ \bibnamefont {Dewhurst}}, \bibinfo {author} {\bibfnamefont
  {G.}~\bibnamefont {Reiss}}, \bibinfo {author} {\bibfnamefont
  {L.}~\bibnamefont {Molenkamp}}, \bibinfo {author} {\bibfnamefont
  {M.}~\bibnamefont {Aeschlimann}}, \bibinfo {author} {\bibfnamefont
  {M.}~\bibnamefont {Cinchetti}}, \bibinfo {author} {\bibfnamefont
  {M.}~\bibnamefont {Münzenberg}}, \bibinfo {author} {\bibfnamefont
  {S.}~\bibnamefont {Sharma}},\ and\ \bibinfo {author} {\bibfnamefont
  {S.}~\bibnamefont {Mathias}},\ }\bibfield  {title} {\enquote {\bibinfo
  {title} {Efficiency of ultrafast optically induced spin transfer in {Heusler}
  compounds},}\ }\href {https://doi.org/10.1103/PhysRevResearch.2.023199}
  {\bibfield  {journal} {\bibinfo  {journal} {Phys. Rev. Research}\ }\textbf
  {\bibinfo {volume} {2}},\ \bibinfo {pages} {023199} (\bibinfo {year}
  {2020})}\BibitemShut {NoStop}%
\bibitem [{\citenamefont {Tengdin}\ \emph {et~al.}(2020)\citenamefont
  {Tengdin}, \citenamefont {Gentry}, \citenamefont {Blonsky}, \citenamefont
  {Zusin}, \citenamefont {Gerrity}, \citenamefont {Hellbrück}, \citenamefont
  {Hofherr}, \citenamefont {Shaw}, \citenamefont {Kvashnin}, \citenamefont
  {Delczeg-Czirjak}, \citenamefont {Arora}, \citenamefont {Nembach},
  \citenamefont {Silva}, \citenamefont {Mathias}, \citenamefont {Aeschlimann},
  \citenamefont {Kapteyn}, \citenamefont {Thonig}, \citenamefont {Koumpouras},
  \citenamefont {Eriksson},\ and\ \citenamefont
  {Murnane}}]{tengdin_direct_2020}%
  \BibitemOpen
  \bibfield  {author} {\bibinfo {author} {\bibfnamefont {P.}~\bibnamefont
  {Tengdin}}, \bibinfo {author} {\bibfnamefont {C.}~\bibnamefont {Gentry}},
  \bibinfo {author} {\bibfnamefont {A.}~\bibnamefont {Blonsky}}, \bibinfo
  {author} {\bibfnamefont {D.}~\bibnamefont {Zusin}}, \bibinfo {author}
  {\bibfnamefont {M.}~\bibnamefont {Gerrity}}, \bibinfo {author} {\bibfnamefont
  {L.}~\bibnamefont {Hellbrück}}, \bibinfo {author} {\bibfnamefont
  {M.}~\bibnamefont {Hofherr}}, \bibinfo {author} {\bibfnamefont
  {J.}~\bibnamefont {Shaw}}, \bibinfo {author} {\bibfnamefont {Y.}~\bibnamefont
  {Kvashnin}}, \bibinfo {author} {\bibfnamefont {E.~K.}\ \bibnamefont
  {Delczeg-Czirjak}}, \bibinfo {author} {\bibfnamefont {M.}~\bibnamefont
  {Arora}}, \bibinfo {author} {\bibfnamefont {H.}~\bibnamefont {Nembach}},
  \bibinfo {author} {\bibfnamefont {T.~J.}\ \bibnamefont {Silva}}, \bibinfo
  {author} {\bibfnamefont {S.}~\bibnamefont {Mathias}}, \bibinfo {author}
  {\bibfnamefont {M.}~\bibnamefont {Aeschlimann}}, \bibinfo {author}
  {\bibfnamefont {H.~C.}\ \bibnamefont {Kapteyn}}, \bibinfo {author}
  {\bibfnamefont {D.}~\bibnamefont {Thonig}}, \bibinfo {author} {\bibfnamefont
  {K.}~\bibnamefont {Koumpouras}}, \bibinfo {author} {\bibfnamefont
  {O.}~\bibnamefont {Eriksson}},\ and\ \bibinfo {author} {\bibfnamefont
  {M.~M.}\ \bibnamefont {Murnane}},\ }\bibfield  {title} {\enquote {\bibinfo
  {title} {Direct light–induced spin transfer between different elements in a
  spintronic {Heusler} material via femtosecond laser excitation},}\ }\href
  {https://doi.org/10.1126/sciadv.aaz1100} {\bibfield  {journal} {\bibinfo
  {journal} {Sci. Adv.}\ }\textbf {\bibinfo {volume} {6}},\ \bibinfo {pages}
  {eaaz1100} (\bibinfo {year} {2020})}\BibitemShut {NoStop}%
\bibitem [{\citenamefont {Hofherr}\ \emph {et~al.}(2020)\citenamefont
  {Hofherr}, \citenamefont {Häuser}, \citenamefont {Dewhurst}, \citenamefont
  {Tengdin}, \citenamefont {Sakshath}, \citenamefont {Nembach}, \citenamefont
  {Weber}, \citenamefont {Shaw}, \citenamefont {Silva}, \citenamefont
  {Kapteyn}, \citenamefont {Cinchetti}, \citenamefont {Rethfeld}, \citenamefont
  {Murnane}, \citenamefont {Steil}, \citenamefont {Stadtmüller}, \citenamefont
  {Sharma}, \citenamefont {Aeschlimann},\ and\ \citenamefont
  {Mathias}}]{hofherr_ultrafast_2020}%
  \BibitemOpen
  \bibfield  {author} {\bibinfo {author} {\bibfnamefont {M.}~\bibnamefont
  {Hofherr}}, \bibinfo {author} {\bibfnamefont {S.}~\bibnamefont {Häuser}},
  \bibinfo {author} {\bibfnamefont {J.~K.}\ \bibnamefont {Dewhurst}}, \bibinfo
  {author} {\bibfnamefont {P.}~\bibnamefont {Tengdin}}, \bibinfo {author}
  {\bibfnamefont {S.}~\bibnamefont {Sakshath}}, \bibinfo {author}
  {\bibfnamefont {H.~T.}\ \bibnamefont {Nembach}}, \bibinfo {author}
  {\bibfnamefont {S.~T.}\ \bibnamefont {Weber}}, \bibinfo {author}
  {\bibfnamefont {J.~M.}\ \bibnamefont {Shaw}}, \bibinfo {author}
  {\bibfnamefont {T.~J.}\ \bibnamefont {Silva}}, \bibinfo {author}
  {\bibfnamefont {H.~C.}\ \bibnamefont {Kapteyn}}, \bibinfo {author}
  {\bibfnamefont {M.}~\bibnamefont {Cinchetti}}, \bibinfo {author}
  {\bibfnamefont {B.}~\bibnamefont {Rethfeld}}, \bibinfo {author}
  {\bibfnamefont {M.~M.}\ \bibnamefont {Murnane}}, \bibinfo {author}
  {\bibfnamefont {D.}~\bibnamefont {Steil}}, \bibinfo {author} {\bibfnamefont
  {B.}~\bibnamefont {Stadtmüller}}, \bibinfo {author} {\bibfnamefont
  {S.}~\bibnamefont {Sharma}}, \bibinfo {author} {\bibfnamefont
  {M.}~\bibnamefont {Aeschlimann}},\ and\ \bibinfo {author} {\bibfnamefont
  {S.}~\bibnamefont {Mathias}},\ }\bibfield  {title} {\enquote {\bibinfo
  {title} {Ultrafast optically induced spin transfer in ferromagnetic
  alloys},}\ }\href {https://doi.org/10.1126/sciadv.aay8717} {\bibfield
  {journal} {\bibinfo  {journal} {Sci. Adv.}\ }\textbf {\bibinfo {volume}
  {6}},\ \bibinfo {pages} {eaay8717} (\bibinfo {year} {2020})}\BibitemShut
  {NoStop}%
\bibitem [{\citenamefont {Willems}\ \emph {et~al.}(2020)\citenamefont
  {Willems}, \citenamefont {von Korff~Schmising}, \citenamefont {Strüber},
  \citenamefont {Schick}, \citenamefont {Engel}, \citenamefont {Dewhurst},
  \citenamefont {Elliott}, \citenamefont {Sharma},\ and\ \citenamefont
  {Eisebitt}}]{willems_optical_2020}%
  \BibitemOpen
  \bibfield  {author} {\bibinfo {author} {\bibfnamefont {F.}~\bibnamefont
  {Willems}}, \bibinfo {author} {\bibfnamefont {C.}~\bibnamefont {von
  Korff~Schmising}}, \bibinfo {author} {\bibfnamefont {C.}~\bibnamefont
  {Strüber}}, \bibinfo {author} {\bibfnamefont {D.}~\bibnamefont {Schick}},
  \bibinfo {author} {\bibfnamefont {D.~W.}\ \bibnamefont {Engel}}, \bibinfo
  {author} {\bibfnamefont {J.~K.}\ \bibnamefont {Dewhurst}}, \bibinfo {author}
  {\bibfnamefont {P.}~\bibnamefont {Elliott}}, \bibinfo {author} {\bibfnamefont
  {S.}~\bibnamefont {Sharma}},\ and\ \bibinfo {author} {\bibfnamefont
  {S.}~\bibnamefont {Eisebitt}},\ }\bibfield  {title} {\enquote {\bibinfo
  {title} {Optical inter-site spin transfer probed by energy and spin-resolved
  transient absorption spectroscopy},}\ }\href
  {https://doi.org/10.1038/s41467-020-14691-5} {\bibfield  {journal} {\bibinfo
  {journal} {Nat. Comm.}\ }\textbf {\bibinfo {volume} {11}},\ \bibinfo {pages}
  {1--7} (\bibinfo {year} {2020})}\BibitemShut {NoStop}%
\bibitem [{\citenamefont {Kawamiya}\ and\ \citenamefont
  {Adachi}(1975)}]{kawamiya_magnetic_1975}%
  \BibitemOpen
  \bibfield  {author} {\bibinfo {author} {\bibfnamefont {N.}~\bibnamefont
  {Kawamiya}}\ and\ \bibinfo {author} {\bibfnamefont {K.}~\bibnamefont
  {Adachi}},\ }\bibfield  {title} {\enquote {\bibinfo {title} {Magnetic
  {Properties} of {Ordered} and {Disordered} {Ni}\_(1-x) {Fe}\_(x) {Pt}},}\
  }\href {https://doi.org/10.2320/matertrans1960.16.327} {\bibfield  {journal}
  {\bibinfo  {journal} {J-Stage}\ }\textbf {\bibinfo {volume} {16}},\ \bibinfo
  {pages} {327--332} (\bibinfo {year} {1975})}\BibitemShut {NoStop}%
\bibitem [{\citenamefont {Dewhurst}\ \emph
  {et~al.}(2018{\natexlab{b}})\citenamefont {Dewhurst}, \citenamefont
  {Shallcross}, \citenamefont {Gross},\ and\ \citenamefont
  {Sharma}}]{dewhurst_substrate-controlled_2018}%
  \BibitemOpen
  \bibfield  {author} {\bibinfo {author} {\bibfnamefont {J.~K.}\ \bibnamefont
  {Dewhurst}}, \bibinfo {author} {\bibfnamefont {S.}~\bibnamefont
  {Shallcross}}, \bibinfo {author} {\bibfnamefont {E.~K.~U.}\ \bibnamefont
  {Gross}},\ and\ \bibinfo {author} {\bibfnamefont {S.}~\bibnamefont
  {Sharma}},\ }\bibfield  {title} {\enquote {\bibinfo {title}
  {Substrate-{Controlled} {Ultrafast} {Spin} {Injection} and
  {Demagnetization}},}\ }\href
  {https://doi.org/10.1103/PhysRevApplied.10.044065} {\bibfield  {journal}
  {\bibinfo  {journal} {Phys. Rev. Appl.}\ }\textbf {\bibinfo {volume} {10}},\
  \bibinfo {pages} {044065} (\bibinfo {year} {2018}{\natexlab{b}})}\BibitemShut
  {NoStop}%
\bibitem [{\citenamefont {Radu}\ \emph {et~al.}(2015)\citenamefont {Radu},
  \citenamefont {Stamm}, \citenamefont {Eschenlohr}, \citenamefont {Radu},
  \citenamefont {Abrudan}, \citenamefont {Vahaplar}, \citenamefont {Kachel},
  \citenamefont {Pontius}, \citenamefont {Mitzner}, \citenamefont {Holldack},
  \citenamefont {Föhlisch}, \citenamefont {Ostler}, \citenamefont {Mentink},
  \citenamefont {Evans}, \citenamefont {Chantrell}, \citenamefont {Tsukamoto},
  \citenamefont {Itoh}, \citenamefont {Kirilyuk}, \citenamefont {Kimel},\ and\
  \citenamefont {Rasing}}]{radu_ultrafast_2015}%
  \BibitemOpen
  \bibfield  {author} {\bibinfo {author} {\bibfnamefont {I.}~\bibnamefont
  {Radu}}, \bibinfo {author} {\bibfnamefont {C.}~\bibnamefont {Stamm}},
  \bibinfo {author} {\bibfnamefont {A.}~\bibnamefont {Eschenlohr}}, \bibinfo
  {author} {\bibfnamefont {F.}~\bibnamefont {Radu}}, \bibinfo {author}
  {\bibfnamefont {R.}~\bibnamefont {Abrudan}}, \bibinfo {author} {\bibfnamefont
  {K.}~\bibnamefont {Vahaplar}}, \bibinfo {author} {\bibfnamefont
  {T.}~\bibnamefont {Kachel}}, \bibinfo {author} {\bibfnamefont
  {N.}~\bibnamefont {Pontius}}, \bibinfo {author} {\bibfnamefont
  {R.}~\bibnamefont {Mitzner}}, \bibinfo {author} {\bibfnamefont
  {K.}~\bibnamefont {Holldack}}, \bibinfo {author} {\bibfnamefont
  {A.}~\bibnamefont {Föhlisch}}, \bibinfo {author} {\bibfnamefont {T.~A.}\
  \bibnamefont {Ostler}}, \bibinfo {author} {\bibfnamefont {J.~H.}\
  \bibnamefont {Mentink}}, \bibinfo {author} {\bibfnamefont {R.~F.~L.}\
  \bibnamefont {Evans}}, \bibinfo {author} {\bibfnamefont {R.~W.}\ \bibnamefont
  {Chantrell}}, \bibinfo {author} {\bibfnamefont {A.}~\bibnamefont
  {Tsukamoto}}, \bibinfo {author} {\bibfnamefont {A.}~\bibnamefont {Itoh}},
  \bibinfo {author} {\bibfnamefont {A.}~\bibnamefont {Kirilyuk}}, \bibinfo
  {author} {\bibfnamefont {A.~V.}\ \bibnamefont {Kimel}},\ and\ \bibinfo
  {author} {\bibfnamefont {T.}~\bibnamefont {Rasing}},\ }\bibfield  {title}
  {\enquote {\bibinfo {title} {Ultrafast and {Distinct} {Spin} {Dynamics} in
  {Magnetic} {Alloys}},}\ }\href {https://doi.org/10.1142/S2010324715500046}
  {\bibfield  {journal} {\bibinfo  {journal} {SPIN}\ }\textbf {\bibinfo
  {volume} {05}},\ \bibinfo {pages} {1550004} (\bibinfo {year}
  {2015})}\BibitemShut {NoStop}%
\bibitem [{\citenamefont {Koopmans}\ \emph {et~al.}(2010)\citenamefont
  {Koopmans}, \citenamefont {Malinowski}, \citenamefont {Dalla~Longa},
  \citenamefont {Steiauf}, \citenamefont {Fähnle}, \citenamefont {Roth},
  \citenamefont {Cinchetti},\ and\ \citenamefont
  {Aeschlimann}}]{koopmans_explaining_2010}%
  \BibitemOpen
  \bibfield  {author} {\bibinfo {author} {\bibfnamefont {B.}~\bibnamefont
  {Koopmans}}, \bibinfo {author} {\bibfnamefont {G.}~\bibnamefont
  {Malinowski}}, \bibinfo {author} {\bibfnamefont {F.}~\bibnamefont
  {Dalla~Longa}}, \bibinfo {author} {\bibfnamefont {D.}~\bibnamefont
  {Steiauf}}, \bibinfo {author} {\bibfnamefont {M.}~\bibnamefont {Fähnle}},
  \bibinfo {author} {\bibfnamefont {T.}~\bibnamefont {Roth}}, \bibinfo {author}
  {\bibfnamefont {M.}~\bibnamefont {Cinchetti}},\ and\ \bibinfo {author}
  {\bibfnamefont {M.}~\bibnamefont {Aeschlimann}},\ }\bibfield  {title}
  {\enquote {\bibinfo {title} {Explaining the paradoxical diversity of
  ultrafast laser-induced demagnetization},}\ }\href
  {https://doi.org/10.1038/nmat2593} {\bibfield  {journal} {\bibinfo  {journal}
  {Nat. Mat.}\ }\textbf {\bibinfo {volume} {9}},\ \bibinfo {pages} {259--265}
  (\bibinfo {year} {2010})}\BibitemShut {NoStop}%
\bibitem [{\citenamefont {Singh}(1994)}]{singh}%
  \BibitemOpen
  \bibfield  {author} {\bibinfo {author} {\bibfnamefont {D.~J.}\ \bibnamefont
  {Singh}},\ }\href@noop {} {\emph {\bibinfo {title} {Planewaves
  Pseudopotentials and the LAPW Method}}}\ (\bibinfo  {publisher} {Kluwer
  Academic Publishers, Boston},\ \bibinfo {year} {1994})\BibitemShut {NoStop}%
\bibitem [{\citenamefont {Dewhurst}, \citenamefont {Sharma},\ and\
  \citenamefont {et~al.}(2018)}]{elk}%
  \BibitemOpen
  \bibfield  {author} {\bibinfo {author} {\bibfnamefont {J.~K.}\ \bibnamefont
  {Dewhurst}}, \bibinfo {author} {\bibfnamefont {S.}~\bibnamefont {Sharma}},\
  and\ \bibinfo {author} {\bibnamefont {et~al.}},\ }\href {elk.sourceforge.net}
  {} (\bibinfo {year} {Jan. 14 {\bf 2018}})\BibitemShut {NoStop}%
\bibitem [{\citenamefont {Dewhurst}\ \emph {et~al.}(2016)\citenamefont
  {Dewhurst}, \citenamefont {Krieger}, \citenamefont {Sharma},\ and\
  \citenamefont {Gross}}]{dewhurst2016}%
  \BibitemOpen
  \bibfield  {author} {\bibinfo {author} {\bibfnamefont {J.~K.}\ \bibnamefont
  {Dewhurst}}, \bibinfo {author} {\bibfnamefont {K.}~\bibnamefont {Krieger}},
  \bibinfo {author} {\bibfnamefont {S.}~\bibnamefont {Sharma}},\ and\ \bibinfo
  {author} {\bibfnamefont {E.~K.~U.}\ \bibnamefont {Gross}},\ }\bibfield
  {title} {\enquote {\bibinfo {title} {An efficient algorithm for time
  propagation as applied to linearized augmented plane wave method},}\ }\href
  {https://doi.org/10.1016/j.cpc.2016.09.001} {\bibfield  {journal} {\bibinfo
  {journal} {Computer Physics Communications}\ }\textbf {\bibinfo {volume}
  {209}},\ \bibinfo {pages} {92--95} (\bibinfo {year} {2016})}\BibitemShut
  {NoStop}%
\bibitem [{\citenamefont {Lebedev}\ \emph {et~al.}(2005)\citenamefont
  {Lebedev}, \citenamefont {Misochko}, \citenamefont {Dekorsy},\ and\
  \citenamefont {Georgiev}}]{Lebedev2005}%
  \BibitemOpen
  \bibfield  {author} {\bibinfo {author} {\bibfnamefont {M.~V.}\ \bibnamefont
  {Lebedev}}, \bibinfo {author} {\bibfnamefont {O.~V.}\ \bibnamefont
  {Misochko}}, \bibinfo {author} {\bibfnamefont {T.}~\bibnamefont {Dekorsy}},\
  and\ \bibinfo {author} {\bibfnamefont {N.}~\bibnamefont {Georgiev}},\
  }\bibfield  {title} {\enquote {\bibinfo {title} {{On the nature of "coherent
  artifact"}},}\ }\href {https://doi.org/10.1134/1.1884668} {\bibfield
  {journal} {\bibinfo  {journal} {J. Exp. Theor. Phys.}\ }\textbf {\bibinfo
  {volume} {100}},\ \bibinfo {pages} {272--282} (\bibinfo {year}
  {2005})}\BibitemShut {NoStop}%
\bibitem [{\citenamefont {Radu}\ \emph {et~al.}(2009)\citenamefont {Radu},
  \citenamefont {Woltersdorf}, \citenamefont {Kiessling}, \citenamefont
  {Melnikov}, \citenamefont {Bovensiepen}, \citenamefont {Thiele},\ and\
  \citenamefont {Back}}]{Radu2009}%
  \BibitemOpen
  \bibfield  {author} {\bibinfo {author} {\bibfnamefont {I.}~\bibnamefont
  {Radu}}, \bibinfo {author} {\bibfnamefont {G.}~\bibnamefont {Woltersdorf}},
  \bibinfo {author} {\bibfnamefont {M.}~\bibnamefont {Kiessling}}, \bibinfo
  {author} {\bibfnamefont {A.}~\bibnamefont {Melnikov}}, \bibinfo {author}
  {\bibfnamefont {U.}~\bibnamefont {Bovensiepen}}, \bibinfo {author}
  {\bibfnamefont {J.-U.}\ \bibnamefont {Thiele}},\ and\ \bibinfo {author}
  {\bibfnamefont {C.~H.}\ \bibnamefont {Back}},\ }\bibfield  {title} {\enquote
  {\bibinfo {title} {{Laser-Induced Magnetization Dynamics of Lanthanide-Doped
  Permalloy Thin Films}},}\ }\href
  {https://doi.org/10.1103/PhysRevLett.102.117201} {\bibfield  {journal}
  {\bibinfo  {journal} {Phys. Rev. Lett.}\ }\textbf {\bibinfo {volume} {102}},\
  \bibinfo {pages} {117201} (\bibinfo {year} {2009})}\BibitemShut {NoStop}%
\bibitem [{\citenamefont {Zhang}\ \emph {et~al.}(2020)\citenamefont {Zhang},
  \citenamefont {Maldonado}, \citenamefont {Jin}, \citenamefont {Seifert},
  \citenamefont {Arabski}, \citenamefont {Schmerber}, \citenamefont
  {Beaurepaire}, \citenamefont {Bonn}, \citenamefont {Kampfrath}, \citenamefont
  {Oppeneer},\ and\ \citenamefont {Turchinovich}}]{Zhang2020}%
  \BibitemOpen
  \bibfield  {author} {\bibinfo {author} {\bibfnamefont {W.}~\bibnamefont
  {Zhang}}, \bibinfo {author} {\bibfnamefont {P.}~\bibnamefont {Maldonado}},
  \bibinfo {author} {\bibfnamefont {Z.}~\bibnamefont {Jin}}, \bibinfo {author}
  {\bibfnamefont {T.~S.}\ \bibnamefont {Seifert}}, \bibinfo {author}
  {\bibfnamefont {J.}~\bibnamefont {Arabski}}, \bibinfo {author} {\bibfnamefont
  {G.}~\bibnamefont {Schmerber}}, \bibinfo {author} {\bibfnamefont
  {E.}~\bibnamefont {Beaurepaire}}, \bibinfo {author} {\bibfnamefont
  {M.}~\bibnamefont {Bonn}}, \bibinfo {author} {\bibfnamefont {T.}~\bibnamefont
  {Kampfrath}}, \bibinfo {author} {\bibfnamefont {P.~M.}\ \bibnamefont
  {Oppeneer}},\ and\ \bibinfo {author} {\bibfnamefont {D.}~\bibnamefont
  {Turchinovich}},\ }\bibfield  {title} {\enquote {\bibinfo {title} {{Ultrafast
  terahertz magnetometry}},}\ }\href
  {https://doi.org/10.1038/s41467-020-17935-6} {\bibfield  {journal} {\bibinfo
  {journal} {Nat. Commun.}\ }\textbf {\bibinfo {volume} {11}},\ \bibinfo
  {pages} {4247} (\bibinfo {year} {2020})}\BibitemShut {NoStop}%
\bibitem [{\citenamefont {Kuiper}\ \emph {et~al.}(2014)\citenamefont {Kuiper},
  \citenamefont {Roth}, \citenamefont {Schellekens}, \citenamefont {Schmitt},
  \citenamefont {Koopmans}, \citenamefont {Cinchetti},\ and\ \citenamefont
  {Aeschlimann}}]{kuiper_spin-orbit_2014}%
  \BibitemOpen
  \bibfield  {author} {\bibinfo {author} {\bibfnamefont {K.~C.}\ \bibnamefont
  {Kuiper}}, \bibinfo {author} {\bibfnamefont {T.}~\bibnamefont {Roth}},
  \bibinfo {author} {\bibfnamefont {A.~J.}\ \bibnamefont {Schellekens}},
  \bibinfo {author} {\bibfnamefont {O.}~\bibnamefont {Schmitt}}, \bibinfo
  {author} {\bibfnamefont {B.}~\bibnamefont {Koopmans}}, \bibinfo {author}
  {\bibfnamefont {M.}~\bibnamefont {Cinchetti}},\ and\ \bibinfo {author}
  {\bibfnamefont {M.}~\bibnamefont {Aeschlimann}},\ }\bibfield  {title}
  {\enquote {\bibinfo {title} {Spin-orbit enhanced demagnetization rate in
  {Co}/{Pt}-multilayers},}\ }\href {https://doi.org/10.1063/1.4902069}
  {\bibfield  {journal} {\bibinfo  {journal} {Appl. Phys. Lett.}\ }\textbf
  {\bibinfo {volume} {105}},\ \bibinfo {pages} {202402} (\bibinfo {year}
  {2014})}\BibitemShut {NoStop}%
\bibitem [{\citenamefont {Eschenlohr}(2012)}]{Eschenlohr2012}%
  \BibitemOpen
  \bibfield  {author} {\bibinfo {author} {\bibfnamefont {A.}~\bibnamefont
  {Eschenlohr}},\ }\emph {\bibinfo {title} {Element-resolved ultrafast
  magnetization dynamics in ferromagnetic alloys and multilayers}},\ \href@noop
  {} {\bibinfo {type} {doctoralthesis}},\ \bibinfo  {school} {Universit{\"a}t
  Potsdam} (\bibinfo {year} {2012})\BibitemShut {NoStop}%
\bibitem [{\citenamefont {Krieger}\ \emph {et~al.}(2015)\citenamefont
  {Krieger}, \citenamefont {Dewhurst}, \citenamefont {Elliott}, \citenamefont
  {Sharma},\ and\ \citenamefont {Gross}}]{krieger2015}%
  \BibitemOpen
  \bibfield  {author} {\bibinfo {author} {\bibfnamefont {K.}~\bibnamefont
  {Krieger}}, \bibinfo {author} {\bibfnamefont {J.~K.}\ \bibnamefont
  {Dewhurst}}, \bibinfo {author} {\bibfnamefont {P.}~\bibnamefont {Elliott}},
  \bibinfo {author} {\bibfnamefont {S.}~\bibnamefont {Sharma}},\ and\ \bibinfo
  {author} {\bibfnamefont {E.~K.~U.}\ \bibnamefont {Gross}},\ }\bibfield
  {title} {\enquote {\bibinfo {title} {Laser-{Induced} {Demagnetization} at
  {Ultrashort} {Time} {Scales}: {Predictions} of {TDDFT}},}\ }\href
  {https://doi.org/10.1021/acs.jctc.5b00621} {\bibfield  {journal} {\bibinfo
  {journal} {Journal of Chemical Theory and Computation}\ }\textbf {\bibinfo
  {volume} {11}},\ \bibinfo {pages} {4870--4874} (\bibinfo {year}
  {2015})}\BibitemShut {NoStop}%
\bibitem [{\citenamefont {Mathias}\ \emph {et~al.}(2012)\citenamefont
  {Mathias}, \citenamefont {La-O-Vorakiat}, \citenamefont {Grychtol},
  \citenamefont {Granitzka}, \citenamefont {Turgut}, \citenamefont {Shaw},
  \citenamefont {Adam}, \citenamefont {Nembach}, \citenamefont {Siemens},
  \citenamefont {Eich}, \citenamefont {Schneider}, \citenamefont {Silva},
  \citenamefont {Aeschlimann}, \citenamefont {Murnane},\ and\ \citenamefont
  {Kapteyn}}]{Mathias2012}%
  \BibitemOpen
  \bibfield  {author} {\bibinfo {author} {\bibfnamefont {S.}~\bibnamefont
  {Mathias}}, \bibinfo {author} {\bibfnamefont {C.}~\bibnamefont
  {La-O-Vorakiat}}, \bibinfo {author} {\bibfnamefont {P.}~\bibnamefont
  {Grychtol}}, \bibinfo {author} {\bibfnamefont {P.}~\bibnamefont {Granitzka}},
  \bibinfo {author} {\bibfnamefont {E.}~\bibnamefont {Turgut}}, \bibinfo
  {author} {\bibfnamefont {J.~M.}\ \bibnamefont {Shaw}}, \bibinfo {author}
  {\bibfnamefont {R.}~\bibnamefont {Adam}}, \bibinfo {author} {\bibfnamefont
  {H.~T.}\ \bibnamefont {Nembach}}, \bibinfo {author} {\bibfnamefont {M.~E.}\
  \bibnamefont {Siemens}}, \bibinfo {author} {\bibfnamefont {S.}~\bibnamefont
  {Eich}}, \bibinfo {author} {\bibfnamefont {C.~M.}\ \bibnamefont {Schneider}},
  \bibinfo {author} {\bibfnamefont {T.~J.}\ \bibnamefont {Silva}}, \bibinfo
  {author} {\bibfnamefont {M.}~\bibnamefont {Aeschlimann}}, \bibinfo {author}
  {\bibfnamefont {M.~M.}\ \bibnamefont {Murnane}},\ and\ \bibinfo {author}
  {\bibfnamefont {H.~C.}\ \bibnamefont {Kapteyn}},\ }\bibfield  {title}
  {\enquote {\bibinfo {title} {{Probing the timescale of the exchange
  interaction in a ferromagnetic alloy}},}\ }\href
  {https://doi.org/10.1073/pnas.1201371109} {\bibfield  {journal} {\bibinfo
  {journal} {Proc. Natl. Acad. Sci.}\ }\textbf {\bibinfo {volume} {109}},\
  \bibinfo {pages} {4792--4797} (\bibinfo {year} {2012})}\BibitemShut {NoStop}%
\bibitem [{\citenamefont {Jana}\ \emph {et~al.}(2017)\citenamefont {Jana},
  \citenamefont {Terschl{\"{u}}sen}, \citenamefont {Stefanuik}, \citenamefont
  {Plogmaker}, \citenamefont {Troisi}, \citenamefont {Malik}, \citenamefont
  {Svanqvist}, \citenamefont {Knut}, \citenamefont {S{\"{o}}derstr{\"{o}}m},\
  and\ \citenamefont {Karis}}]{Jana2017}%
  \BibitemOpen
  \bibfield  {author} {\bibinfo {author} {\bibfnamefont {S.}~\bibnamefont
  {Jana}}, \bibinfo {author} {\bibfnamefont {J.~A.}\ \bibnamefont
  {Terschl{\"{u}}sen}}, \bibinfo {author} {\bibfnamefont {R.}~\bibnamefont
  {Stefanuik}}, \bibinfo {author} {\bibfnamefont {S.}~\bibnamefont
  {Plogmaker}}, \bibinfo {author} {\bibfnamefont {S.}~\bibnamefont {Troisi}},
  \bibinfo {author} {\bibfnamefont {R.~S.}\ \bibnamefont {Malik}}, \bibinfo
  {author} {\bibfnamefont {M.}~\bibnamefont {Svanqvist}}, \bibinfo {author}
  {\bibfnamefont {R.}~\bibnamefont {Knut}}, \bibinfo {author} {\bibfnamefont
  {J.}~\bibnamefont {S{\"{o}}derstr{\"{o}}m}},\ and\ \bibinfo {author}
  {\bibfnamefont {O.}~\bibnamefont {Karis}},\ }\bibfield  {title} {\enquote
  {\bibinfo {title} {{A setup for element specific magnetization dynamics using
  the transverse magneto-optic Kerr effect in the energy range of 30-72 eV}},}\
  }\href {https://doi.org/10.1063/1.4978907} {\bibfield  {journal} {\bibinfo
  {journal} {Rev. Sci. Instrum.}\ }\textbf {\bibinfo {volume} {88}},\ \bibinfo
  {pages} {033113} (\bibinfo {year} {2017})}\BibitemShut {NoStop}%
\end{thebibliography}%


\begin{thebibliography}{13}%
\makeatletter
\providecommand \@ifxundefined [1]{%
 \@ifx{#1\undefined}
}%
\providecommand \@ifnum [1]{%
 \ifnum #1\expandafter \@firstoftwo
 \else \expandafter \@secondoftwo
 \fi
}%
\providecommand \@ifx [1]{%
 \ifx #1\expandafter \@firstoftwo
 \else \expandafter \@secondoftwo
 \fi
}%
\providecommand \natexlab [1]{#1}%
\providecommand \enquote  [1]{``#1''}%
\providecommand \bibnamefont  [1]{#1}%
\providecommand \bibfnamefont [1]{#1}%
\providecommand \citenamefont [1]{#1}%
\providecommand \href@noop [0]{\@secondoftwo}%
\providecommand \href [0]{\begingroup \@sanitize@url \@href}%
\providecommand \@href[1]{\@@startlink{#1}\@@href}%
\providecommand \@@href[1]{\endgroup#1\@@endlink}%
\providecommand \@sanitize@url [0]{\catcode `\\12\catcode `\$12\catcode
  `\&12\catcode `\#12\catcode `\^12\catcode `\_12\catcode `\%12\relax}%
\providecommand \@@startlink[1]{}%
\providecommand \@@endlink[0]{}%
\providecommand \url  [0]{\begingroup\@sanitize@url \@url }%
\providecommand \@url [1]{\endgroup\@href {#1}{\urlprefix }}%
\providecommand \urlprefix  [0]{URL }%
\providecommand \Eprint [0]{\href }%
\providecommand \doibase [0]{https://doi.org/}%
\providecommand \selectlanguage [0]{\@gobble}%
\providecommand \bibinfo  [0]{\@secondoftwo}%
\providecommand \bibfield  [0]{\@secondoftwo}%
\providecommand \translation [1]{[#1]}%
\providecommand \BibitemOpen [0]{}%
\providecommand \bibitemStop [0]{}%
\providecommand \bibitemNoStop [0]{.\EOS\space}%
\providecommand \EOS [0]{\spacefactor3000\relax}%
\providecommand \BibitemShut  [1]{\csname bibitem#1\endcsname}%
\let\auto@bib@innerbib\@empty
\bibitem [{\citenamefont {Lebedev}\ \emph {et~al.}(2005)\citenamefont
  {Lebedev}, \citenamefont {Misochko}, \citenamefont {Dekorsy},\ and\
  \citenamefont {Georgiev}}]{Lebedev2005}%
  \BibitemOpen
  \bibfield  {author} {\bibinfo {author} {\bibfnamefont {M.~V.}\ \bibnamefont
  {Lebedev}}, \bibinfo {author} {\bibfnamefont {O.~V.}\ \bibnamefont
  {Misochko}}, \bibinfo {author} {\bibfnamefont {T.}~\bibnamefont {Dekorsy}},\
  and\ \bibinfo {author} {\bibfnamefont {N.}~\bibnamefont {Georgiev}},\
  }\bibfield  {title} {\enquote {\bibinfo {title} {{On the nature of "coherent
  artifact"}},}\ }\href {https://doi.org/10.1134/1.1884668} {\bibfield
  {journal} {\bibinfo  {journal} {J. Exp. Theor. Phys.}\ }\textbf {\bibinfo
  {volume} {100}},\ \bibinfo {pages} {272--282} (\bibinfo {year}
  {2005})}\BibitemShut {NoStop}%
\bibitem [{\citenamefont {Radu}\ \emph {et~al.}(2009)\citenamefont {Radu},
  \citenamefont {Woltersdorf}, \citenamefont {Kiessling}, \citenamefont
  {Melnikov}, \citenamefont {Bovensiepen}, \citenamefont {Thiele},\ and\
  \citenamefont {Back}}]{Radu2009}%
  \BibitemOpen
  \bibfield  {author} {\bibinfo {author} {\bibfnamefont {I.}~\bibnamefont
  {Radu}}, \bibinfo {author} {\bibfnamefont {G.}~\bibnamefont {Woltersdorf}},
  \bibinfo {author} {\bibfnamefont {M.}~\bibnamefont {Kiessling}}, \bibinfo
  {author} {\bibfnamefont {A.}~\bibnamefont {Melnikov}}, \bibinfo {author}
  {\bibfnamefont {U.}~\bibnamefont {Bovensiepen}}, \bibinfo {author}
  {\bibfnamefont {J.-U.}\ \bibnamefont {Thiele}},\ and\ \bibinfo {author}
  {\bibfnamefont {C.~H.}\ \bibnamefont {Back}},\ }\bibfield  {title} {\enquote
  {\bibinfo {title} {{Laser-Induced Magnetization Dynamics of Lanthanide-Doped
  Permalloy Thin Films}},}\ }\href
  {https://doi.org/10.1103/PhysRevLett.102.117201} {\bibfield  {journal}
  {\bibinfo  {journal} {Phys. Rev. Lett.}\ }\textbf {\bibinfo {volume} {102}},\
  \bibinfo {pages} {117201} (\bibinfo {year} {2009})}\BibitemShut {NoStop}%
\bibitem [{\citenamefont {Dewhurst}\ \emph {et~al.}(2018)\citenamefont
  {Dewhurst}, \citenamefont {Shallcross}, \citenamefont {Gross},\ and\
  \citenamefont {Sharma}}]{dewhurst_substrate-controlled_2018}%
  \BibitemOpen
  \bibfield  {author} {\bibinfo {author} {\bibfnamefont {J.~K.}\ \bibnamefont
  {Dewhurst}}, \bibinfo {author} {\bibfnamefont {S.}~\bibnamefont
  {Shallcross}}, \bibinfo {author} {\bibfnamefont {E.~K.~U.}\ \bibnamefont
  {Gross}},\ and\ \bibinfo {author} {\bibfnamefont {S.}~\bibnamefont
  {Sharma}},\ }\bibfield  {title} {\enquote {\bibinfo {title}
  {Substrate-{Controlled} {Ultrafast} {Spin} {Injection} and
  {Demagnetization}},}\ }\href
  {https://doi.org/10.1103/PhysRevApplied.10.044065} {\bibfield  {journal}
  {\bibinfo  {journal} {Phys. Rev. Appl.}\ }\textbf {\bibinfo {volume} {10}},\
  \bibinfo {pages} {044065} (\bibinfo {year} {2018})}\BibitemShut {NoStop}%
\bibitem [{\citenamefont {Zhang}\ \emph {et~al.}(2020)\citenamefont {Zhang},
  \citenamefont {Maldonado}, \citenamefont {Jin}, \citenamefont {Seifert},
  \citenamefont {Arabski}, \citenamefont {Schmerber}, \citenamefont
  {Beaurepaire}, \citenamefont {Bonn}, \citenamefont {Kampfrath}, \citenamefont
  {Oppeneer},\ and\ \citenamefont {Turchinovich}}]{Zhang2020}%
  \BibitemOpen
  \bibfield  {author} {\bibinfo {author} {\bibfnamefont {W.}~\bibnamefont
  {Zhang}}, \bibinfo {author} {\bibfnamefont {P.}~\bibnamefont {Maldonado}},
  \bibinfo {author} {\bibfnamefont {Z.}~\bibnamefont {Jin}}, \bibinfo {author}
  {\bibfnamefont {T.~S.}\ \bibnamefont {Seifert}}, \bibinfo {author}
  {\bibfnamefont {J.}~\bibnamefont {Arabski}}, \bibinfo {author} {\bibfnamefont
  {G.}~\bibnamefont {Schmerber}}, \bibinfo {author} {\bibfnamefont
  {E.}~\bibnamefont {Beaurepaire}}, \bibinfo {author} {\bibfnamefont
  {M.}~\bibnamefont {Bonn}}, \bibinfo {author} {\bibfnamefont {T.}~\bibnamefont
  {Kampfrath}}, \bibinfo {author} {\bibfnamefont {P.~M.}\ \bibnamefont
  {Oppeneer}},\ and\ \bibinfo {author} {\bibfnamefont {D.}~\bibnamefont
  {Turchinovich}},\ }\bibfield  {title} {\enquote {\bibinfo {title} {{Ultrafast
  terahertz magnetometry}},}\ }\href
  {https://doi.org/10.1038/s41467-020-17935-6} {\bibfield  {journal} {\bibinfo
  {journal} {Nat. Commun.}\ }\textbf {\bibinfo {volume} {11}},\ \bibinfo
  {pages} {4247} (\bibinfo {year} {2020})}\BibitemShut {NoStop}%
\bibitem [{\citenamefont {Radu}\ \emph {et~al.}(2015)\citenamefont {Radu},
  \citenamefont {Stamm}, \citenamefont {Eschenlohr}, \citenamefont {Radu},
  \citenamefont {Abrudan}, \citenamefont {Vahaplar}, \citenamefont {Kachel},
  \citenamefont {Pontius}, \citenamefont {Mitzner}, \citenamefont {Holldack},
  \citenamefont {Föhlisch}, \citenamefont {Ostler}, \citenamefont {Mentink},
  \citenamefont {Evans}, \citenamefont {Chantrell}, \citenamefont {Tsukamoto},
  \citenamefont {Itoh}, \citenamefont {Kirilyuk}, \citenamefont {Kimel},\ and\
  \citenamefont {Rasing}}]{radu_ultrafast_2015}%
  \BibitemOpen
  \bibfield  {author} {\bibinfo {author} {\bibfnamefont {I.}~\bibnamefont
  {Radu}}, \bibinfo {author} {\bibfnamefont {C.}~\bibnamefont {Stamm}},
  \bibinfo {author} {\bibfnamefont {A.}~\bibnamefont {Eschenlohr}}, \bibinfo
  {author} {\bibfnamefont {F.}~\bibnamefont {Radu}}, \bibinfo {author}
  {\bibfnamefont {R.}~\bibnamefont {Abrudan}}, \bibinfo {author} {\bibfnamefont
  {K.}~\bibnamefont {Vahaplar}}, \bibinfo {author} {\bibfnamefont
  {T.}~\bibnamefont {Kachel}}, \bibinfo {author} {\bibfnamefont
  {N.}~\bibnamefont {Pontius}}, \bibinfo {author} {\bibfnamefont
  {R.}~\bibnamefont {Mitzner}}, \bibinfo {author} {\bibfnamefont
  {K.}~\bibnamefont {Holldack}}, \bibinfo {author} {\bibfnamefont
  {A.}~\bibnamefont {Föhlisch}}, \bibinfo {author} {\bibfnamefont {T.~A.}\
  \bibnamefont {Ostler}}, \bibinfo {author} {\bibfnamefont {J.~H.}\
  \bibnamefont {Mentink}}, \bibinfo {author} {\bibfnamefont {R.~F.~L.}\
  \bibnamefont {Evans}}, \bibinfo {author} {\bibfnamefont {R.~W.}\ \bibnamefont
  {Chantrell}}, \bibinfo {author} {\bibfnamefont {A.}~\bibnamefont
  {Tsukamoto}}, \bibinfo {author} {\bibfnamefont {A.}~\bibnamefont {Itoh}},
  \bibinfo {author} {\bibfnamefont {A.}~\bibnamefont {Kirilyuk}}, \bibinfo
  {author} {\bibfnamefont {A.~V.}\ \bibnamefont {Kimel}},\ and\ \bibinfo
  {author} {\bibfnamefont {T.}~\bibnamefont {Rasing}},\ }\bibfield  {title}
  {\enquote {\bibinfo {title} {Ultrafast and {Distinct} {Spin} {Dynamics} in
  {Magnetic} {Alloys}},}\ }\href {https://doi.org/10.1142/S2010324715500046}
  {\bibfield  {journal} {\bibinfo  {journal} {SPIN}\ }\textbf {\bibinfo
  {volume} {05}},\ \bibinfo {pages} {1550004} (\bibinfo {year}
  {2015})}\BibitemShut {NoStop}%
\bibitem [{\citenamefont {Koopmans}\ \emph {et~al.}(2010)\citenamefont
  {Koopmans}, \citenamefont {Malinowski}, \citenamefont {Dalla~Longa},
  \citenamefont {Steiauf}, \citenamefont {Fähnle}, \citenamefont {Roth},
  \citenamefont {Cinchetti},\ and\ \citenamefont
  {Aeschlimann}}]{koopmans_explaining_2010}%
  \BibitemOpen
  \bibfield  {author} {\bibinfo {author} {\bibfnamefont {B.}~\bibnamefont
  {Koopmans}}, \bibinfo {author} {\bibfnamefont {G.}~\bibnamefont
  {Malinowski}}, \bibinfo {author} {\bibfnamefont {F.}~\bibnamefont
  {Dalla~Longa}}, \bibinfo {author} {\bibfnamefont {D.}~\bibnamefont
  {Steiauf}}, \bibinfo {author} {\bibfnamefont {M.}~\bibnamefont {Fähnle}},
  \bibinfo {author} {\bibfnamefont {T.}~\bibnamefont {Roth}}, \bibinfo {author}
  {\bibfnamefont {M.}~\bibnamefont {Cinchetti}},\ and\ \bibinfo {author}
  {\bibfnamefont {M.}~\bibnamefont {Aeschlimann}},\ }\bibfield  {title}
  {\enquote {\bibinfo {title} {Explaining the paradoxical diversity of
  ultrafast laser-induced demagnetization},}\ }\href
  {https://doi.org/10.1038/nmat2593} {\bibfield  {journal} {\bibinfo  {journal}
  {Nat. Mat.}\ }\textbf {\bibinfo {volume} {9}},\ \bibinfo {pages} {259--265}
  (\bibinfo {year} {2010})}\BibitemShut {NoStop}%
\bibitem [{\citenamefont {Kuiper}\ \emph {et~al.}(2014)\citenamefont {Kuiper},
  \citenamefont {Roth}, \citenamefont {Schellekens}, \citenamefont {Schmitt},
  \citenamefont {Koopmans}, \citenamefont {Cinchetti},\ and\ \citenamefont
  {Aeschlimann}}]{kuiper_spin-orbit_2014}%
  \BibitemOpen
  \bibfield  {author} {\bibinfo {author} {\bibfnamefont {K.~C.}\ \bibnamefont
  {Kuiper}}, \bibinfo {author} {\bibfnamefont {T.}~\bibnamefont {Roth}},
  \bibinfo {author} {\bibfnamefont {A.~J.}\ \bibnamefont {Schellekens}},
  \bibinfo {author} {\bibfnamefont {O.}~\bibnamefont {Schmitt}}, \bibinfo
  {author} {\bibfnamefont {B.}~\bibnamefont {Koopmans}}, \bibinfo {author}
  {\bibfnamefont {M.}~\bibnamefont {Cinchetti}},\ and\ \bibinfo {author}
  {\bibfnamefont {M.}~\bibnamefont {Aeschlimann}},\ }\bibfield  {title}
  {\enquote {\bibinfo {title} {Spin-orbit enhanced demagnetization rate in
  {Co}/{Pt}-multilayers},}\ }\href {https://doi.org/10.1063/1.4902069}
  {\bibfield  {journal} {\bibinfo  {journal} {Appl. Phys. Lett.}\ }\textbf
  {\bibinfo {volume} {105}},\ \bibinfo {pages} {202402} (\bibinfo {year}
  {2014})}\BibitemShut {NoStop}%
\bibitem [{\citenamefont {Willems}\ \emph {et~al.}(2020)\citenamefont
  {Willems}, \citenamefont {von Korff~Schmising}, \citenamefont {Strüber},
  \citenamefont {Schick}, \citenamefont {Engel}, \citenamefont {Dewhurst},
  \citenamefont {Elliott}, \citenamefont {Sharma},\ and\ \citenamefont
  {Eisebitt}}]{willems_optical_2020}%
  \BibitemOpen
  \bibfield  {author} {\bibinfo {author} {\bibfnamefont {F.}~\bibnamefont
  {Willems}}, \bibinfo {author} {\bibfnamefont {C.}~\bibnamefont {von
  Korff~Schmising}}, \bibinfo {author} {\bibfnamefont {C.}~\bibnamefont
  {Strüber}}, \bibinfo {author} {\bibfnamefont {D.}~\bibnamefont {Schick}},
  \bibinfo {author} {\bibfnamefont {D.~W.}\ \bibnamefont {Engel}}, \bibinfo
  {author} {\bibfnamefont {J.~K.}\ \bibnamefont {Dewhurst}}, \bibinfo {author}
  {\bibfnamefont {P.}~\bibnamefont {Elliott}}, \bibinfo {author} {\bibfnamefont
  {S.}~\bibnamefont {Sharma}},\ and\ \bibinfo {author} {\bibfnamefont
  {S.}~\bibnamefont {Eisebitt}},\ }\bibfield  {title} {\enquote {\bibinfo
  {title} {Optical inter-site spin transfer probed by energy and spin-resolved
  transient absorption spectroscopy},}\ }\href
  {https://doi.org/10.1038/s41467-020-14691-5} {\bibfield  {journal} {\bibinfo
  {journal} {Nat. Comm.}\ }\textbf {\bibinfo {volume} {11}},\ \bibinfo {pages}
  {1--7} (\bibinfo {year} {2020})}\BibitemShut {NoStop}%
\bibitem [{\citenamefont {Eschenlohr}(2012)}]{Eschenlohr2012}%
  \BibitemOpen
  \bibfield  {author} {\bibinfo {author} {\bibfnamefont {A.}~\bibnamefont
  {Eschenlohr}},\ }\emph {\bibinfo {title} {Element-resolved ultrafast
  magnetization dynamics in ferromagnetic alloys and multilayers}},\ \href@noop
  {} {\bibinfo {type} {doctoralthesis}},\ \bibinfo  {school} {Universit{\"a}t
  Potsdam} (\bibinfo {year} {2012})\BibitemShut {NoStop}%
\bibitem [{\citenamefont {Krieger}\ \emph {et~al.}(2015)\citenamefont
  {Krieger}, \citenamefont {Dewhurst}, \citenamefont {Elliott}, \citenamefont
  {Sharma},\ and\ \citenamefont {Gross}}]{krieger2015}%
  \BibitemOpen
  \bibfield  {author} {\bibinfo {author} {\bibfnamefont {K.}~\bibnamefont
  {Krieger}}, \bibinfo {author} {\bibfnamefont {J.~K.}\ \bibnamefont
  {Dewhurst}}, \bibinfo {author} {\bibfnamefont {P.}~\bibnamefont {Elliott}},
  \bibinfo {author} {\bibfnamefont {S.}~\bibnamefont {Sharma}},\ and\ \bibinfo
  {author} {\bibfnamefont {E.~K.~U.}\ \bibnamefont {Gross}},\ }\bibfield
  {title} {\enquote {\bibinfo {title} {Laser-{Induced} {Demagnetization} at
  {Ultrashort} {Time} {Scales}: {Predictions} of {TDDFT}},}\ }\href
  {https://doi.org/10.1021/acs.jctc.5b00621} {\bibfield  {journal} {\bibinfo
  {journal} {Journal of Chemical Theory and Computation}\ }\textbf {\bibinfo
  {volume} {11}},\ \bibinfo {pages} {4870--4874} (\bibinfo {year}
  {2015})}\BibitemShut {NoStop}%
\bibitem [{\citenamefont {Dewhurst}\ \emph {et~al.}(2016)\citenamefont
  {Dewhurst}, \citenamefont {Krieger}, \citenamefont {Sharma},\ and\
  \citenamefont {Gross}}]{dewhurst2016}%
  \BibitemOpen
  \bibfield  {author} {\bibinfo {author} {\bibfnamefont {J.~K.}\ \bibnamefont
  {Dewhurst}}, \bibinfo {author} {\bibfnamefont {K.}~\bibnamefont {Krieger}},
  \bibinfo {author} {\bibfnamefont {S.}~\bibnamefont {Sharma}},\ and\ \bibinfo
  {author} {\bibfnamefont {E.~K.~U.}\ \bibnamefont {Gross}},\ }\bibfield
  {title} {\enquote {\bibinfo {title} {An efficient algorithm for time
  propagation as applied to linearized augmented plane wave method},}\ }\href
  {https://doi.org/10.1016/j.cpc.2016.09.001} {\bibfield  {journal} {\bibinfo
  {journal} {Computer Physics Communications}\ }\textbf {\bibinfo {volume}
  {209}},\ \bibinfo {pages} {92--95} (\bibinfo {year} {2016})}\BibitemShut
  {NoStop}%
\bibitem [{\citenamefont {Mathias}\ \emph {et~al.}(2012)\citenamefont
  {Mathias}, \citenamefont {La-O-Vorakiat}, \citenamefont {Grychtol},
  \citenamefont {Granitzka}, \citenamefont {Turgut}, \citenamefont {Shaw},
  \citenamefont {Adam}, \citenamefont {Nembach}, \citenamefont {Siemens},
  \citenamefont {Eich}, \citenamefont {Schneider}, \citenamefont {Silva},
  \citenamefont {Aeschlimann}, \citenamefont {Murnane},\ and\ \citenamefont
  {Kapteyn}}]{Mathias2012}%
  \BibitemOpen
  \bibfield  {author} {\bibinfo {author} {\bibfnamefont {S.}~\bibnamefont
  {Mathias}}, \bibinfo {author} {\bibfnamefont {C.}~\bibnamefont
  {La-O-Vorakiat}}, \bibinfo {author} {\bibfnamefont {P.}~\bibnamefont
  {Grychtol}}, \bibinfo {author} {\bibfnamefont {P.}~\bibnamefont {Granitzka}},
  \bibinfo {author} {\bibfnamefont {E.}~\bibnamefont {Turgut}}, \bibinfo
  {author} {\bibfnamefont {J.~M.}\ \bibnamefont {Shaw}}, \bibinfo {author}
  {\bibfnamefont {R.}~\bibnamefont {Adam}}, \bibinfo {author} {\bibfnamefont
  {H.~T.}\ \bibnamefont {Nembach}}, \bibinfo {author} {\bibfnamefont {M.~E.}\
  \bibnamefont {Siemens}}, \bibinfo {author} {\bibfnamefont {S.}~\bibnamefont
  {Eich}}, \bibinfo {author} {\bibfnamefont {C.~M.}\ \bibnamefont {Schneider}},
  \bibinfo {author} {\bibfnamefont {T.~J.}\ \bibnamefont {Silva}}, \bibinfo
  {author} {\bibfnamefont {M.}~\bibnamefont {Aeschlimann}}, \bibinfo {author}
  {\bibfnamefont {M.~M.}\ \bibnamefont {Murnane}},\ and\ \bibinfo {author}
  {\bibfnamefont {H.~C.}\ \bibnamefont {Kapteyn}},\ }\bibfield  {title}
  {\enquote {\bibinfo {title} {{Probing the timescale of the exchange
  interaction in a ferromagnetic alloy}},}\ }\href
  {https://doi.org/10.1073/pnas.1201371109} {\bibfield  {journal} {\bibinfo
  {journal} {Proc. Natl. Acad. Sci.}\ }\textbf {\bibinfo {volume} {109}},\
  \bibinfo {pages} {4792--4797} (\bibinfo {year} {2012})}\BibitemShut {NoStop}%
\bibitem [{\citenamefont {Jana}\ \emph {et~al.}(2017)\citenamefont {Jana},
  \citenamefont {Terschl{\"{u}}sen}, \citenamefont {Stefanuik}, \citenamefont
  {Plogmaker}, \citenamefont {Troisi}, \citenamefont {Malik}, \citenamefont
  {Svanqvist}, \citenamefont {Knut}, \citenamefont {S{\"{o}}derstr{\"{o}}m},\
  and\ \citenamefont {Karis}}]{Jana2017}%
  \BibitemOpen
  \bibfield  {author} {\bibinfo {author} {\bibfnamefont {S.}~\bibnamefont
  {Jana}}, \bibinfo {author} {\bibfnamefont {J.~A.}\ \bibnamefont
  {Terschl{\"{u}}sen}}, \bibinfo {author} {\bibfnamefont {R.}~\bibnamefont
  {Stefanuik}}, \bibinfo {author} {\bibfnamefont {S.}~\bibnamefont
  {Plogmaker}}, \bibinfo {author} {\bibfnamefont {S.}~\bibnamefont {Troisi}},
  \bibinfo {author} {\bibfnamefont {R.~S.}\ \bibnamefont {Malik}}, \bibinfo
  {author} {\bibfnamefont {M.}~\bibnamefont {Svanqvist}}, \bibinfo {author}
  {\bibfnamefont {R.}~\bibnamefont {Knut}}, \bibinfo {author} {\bibfnamefont
  {J.}~\bibnamefont {S{\"{o}}derstr{\"{o}}m}},\ and\ \bibinfo {author}
  {\bibfnamefont {O.}~\bibnamefont {Karis}},\ }\bibfield  {title} {\enquote
  {\bibinfo {title} {{A setup for element specific magnetization dynamics using
  the transverse magneto-optic Kerr effect in the energy range of 30-72 eV}},}\
  }\href {https://doi.org/10.1063/1.4978907} {\bibfield  {journal} {\bibinfo
  {journal} {Rev. Sci. Instrum.}\ }\textbf {\bibinfo {volume} {88}},\ \bibinfo
  {pages} {033113} (\bibinfo {year} {2017})}\BibitemShut {NoStop}%
\end{thebibliography}%

\end{document}


\preprint{AIP/123-QED}

    \renewcommand{\textfraction}{0.0}
    \renewcommand{\topfraction}{1.0}
    \renewcommand{\bottomfraction}{1.0}

    \renewcommand{\v}[1]{\bm{\mathrm{#1}}}
    \newcommand{\m}[1]{\bm{\mathsf{#1}}}
    \newcommand{\tx}[1]{\text{#1}}

\title{Supplementary Information: Uncovering the role of the density of states in controlling ultrafast spin dynamics}

\author{Martin Borchert}
\author{Clemens von Korff Schmising}%
\author{Daniel Schick}%
\author{Dieter Engel}%
\author{Sangeeta Sharma}%
\author{Sam Shallcross}
\affiliation{Max-Born-Institut f\"ur Nichtlineare Optik und Kurzzeitspektroskopie, 12489 Berlin, Germany}
\author{Stefan Eisebitt}%
\affiliation{Max-Born-Institut f\"ur Nichtlineare Optik und Kurzzeitspektroskopie, 12489 Berlin, Germany}
\affiliation{Technische Universit\"at Berlin, Institut f\"ur Optik und Atomare Physik, Strasse des 17. Juni 135, 10623 Berlin }

\maketitle

\section{Experimental Setup}

\begin{figure}
	\centering
	\includegraphics[width=1\linewidth]{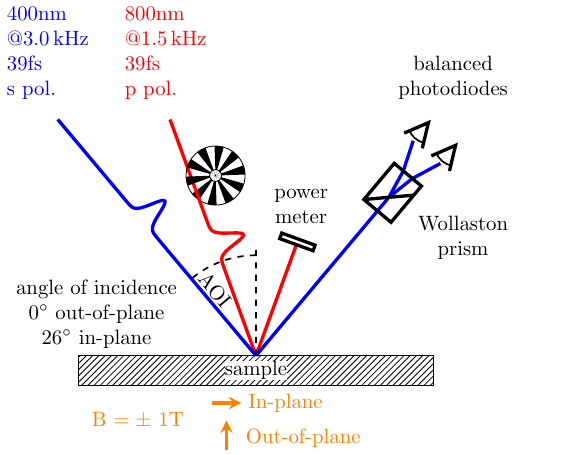}
	\caption{Setup for measuring the time dependent magnetization, $M(t)$, after optical excitation via the magneto optical Kerr effect (MOKE). Changes of the polarization of the incident optical probe pulse ($\lambda_\mathrm{pump} = 400$\,nm) upon reflection off a magnetic sample is detected by balanced photo diodes as a relative intensity change after transmission through a Wollaston prism, yielding a signal proportional to $M(t)$. The pump pulses ($\lambda_\mathrm{pump} = 800$\,nm) arrive at the sample at a variable time delay. The incident, reflected, and transmitted (not shown) pump power is measured during the experiment. The sample can be magnetized in-plane (Longitudinal-MOKE) and out-of-plane (Polar-MOKE) with an external magnetic field of up to 1\,T.}
	\label{fig:MOKE}
\end{figure}

The MOKE describes changes to light upon reflection off a magnetized material. In a polar (P-MOKE) or longitudinal (L-MOKE) geometry the polarization plane of linearly-polarized light is, upon reflection, rotated and made partially elliptical. Both effects are proportional to the component of the magnetization parallel to the propagation direction of the incident light. A schematic of our setup is shown in Fig.~\ref{fig:MOKE}.

The sensitivity of MOKE measurements relies on a high degree of linear polarization of the probe pulse, which we achieve by a set of four reflection polarizers. With a pulse duration of 39\,fs (FWHM) of the pump  ($\lambda_\mathrm{pump} = 800$\,nm) and probe ($\lambda_\mathrm{probe} = 400$\,nm) pulses, we measure a corresponding cross-correlation between the two pulses of $\sigma=55$\,fs, defining the overall temporal resolution of the experiment. To conserve this high time resolution, pump and probe beams are kept colinear to within a few degrees and transmissive optics are avoided whenever possible.

By rotating a $\lambda/2$ wave plate, which sets the \textit{s}- and \textit{p}-components of the light in front of the subsequent reflective \textit{p}-polarizer, we vary the incident fluence on the sample between 0.5\,mJ/cm$^2$ and 20\,mJ/cm$^2$. We determine the absolute values of the absorbed fluence by careful measurements of the pump spot footprints (for different angles of incidence (AOI$\approx26^\circ$ for L-MOKE and $\approx0^\circ$ for P-MOKE) at the sample position with a calibrated beam profiler camera and by determining the transmitted (a few percent) and the reflected (50-70\%) pump power.

After reflection off the sample, the probe beam is split into its \textit{s}- and \textit{p}-components by a Wollaston prism and their differential intensity is detected by a balanced photo diode. This signal is, to first order, directly proportional to the Kerr rotation and therefore to the magnetization.

To determine the absolute value of the laser-induced magnetization change for each time delay between the pump and probe pulses, we toggle the magnetic field between the saturation field $B_\text{sat}$ and $-B_\text{sat}$ of the respective sample and measure the corresponding transient MOKE signal at a 3\,kHz repetition rate. The difference between these two signals is proportional to the transient magnetization $M(t)$. By following this procedure, we also subtract out any non-magnetic contributions from the signal. Together with orthogonal polarizations and non-degenerate photon energies of pump and probe beams, we also avoid any signal contributions from coherent artifacts \cite{Lebedev2005,Radu2009}. To increase the signal-to-noise ratio, for every other probe pulse we block the pump pulse with a mechanical chopper and record the signal by a lock-in-amplifier-based, digital box-car integrator. Finally, we normalize our data to unity for delay points before time zero, yielding the relevant observable $M(t)/M_\mathrm{0}$, where $M_\mathrm{0} = M(t<0)$.

A high degree of automation of the relevant optomechanical components, e.g. the $\lambda$/2 wave plate to set the incident fluence, the mirrors that control the spatial pump-probe overlap and the delay stage, allow for continuous, systematic measurements for a large number of different parameters without human interference.

\section{Comparison of the Ultrafast Dynamics of Alloys and Multilayers}

\begin{figure}
	\includegraphics[width=0.9\linewidth]{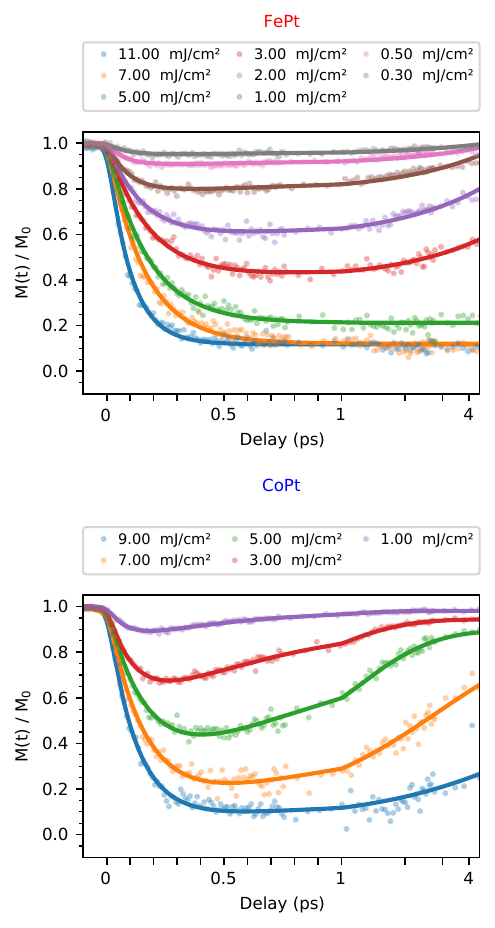}
	\caption{Normalized magnetization $M(t)/M_\mathrm{0}$ for the FMPt alloys for different \textit{incident} fluences, measured in an in-plane geometry. Note that the time axis is linear up to 1\,ps and logarithmic for $t>1$~ps. Solid lines are fits with a double exponential function according to Eq.~1 in the main text.}
	\label{fig:rawalloys}
\end{figure}

In Fig.~\ref{fig:rawalloys}, we display the ultrafast magnetization dynamics of the CoPt and FePt alloys. Note that the time axis is linear up to 1\,ps and logarithmic for $t>1$~ps. We extract the demagnetization amplitudes, $A$, by fitting the data with the double exponential equation shown in the main text (Eq.~1). 

\begin{figure}
	\centering
	\includegraphics[width=0.6\linewidth]{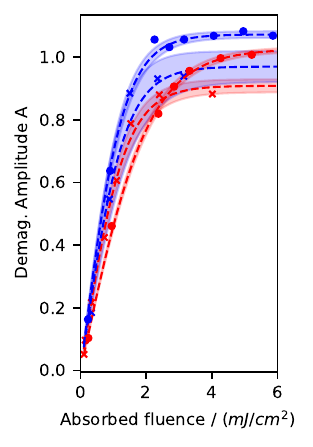}
	\caption{Comparison of demagnetization amplitudes of the FePt and CoPt alloys and Fe$\vert$Pt and Co$\vert$Pt ML for different \textit{absorbed} fluences. Solid lines represent a fit of the data according to a linear-logistic function. The shaded area represents one standard deviation of the fit.}
	\label{MLA}
\end{figure}

The comparison of the demagnetization amplitudes, $A$, for the alloy and multilayer systems are displayed in Fig.~\ref{MLA}. While these systems are of course structurally distinct, their demagnetization amplitudes as a function of fluence are very similar. This suggests that the demagnetization of both systems is dominated by the same underlying mechanism, which -- importantly -- is in line with the OISTR mechanism, that has been shown to allow both intersite (alloy) and interface (ML) spin transfer \cite{dewhurst_substrate-controlled_2018}. 

\section{Oscillations of Cobalt magnetization}

The oscillations of the magnetization of Co seen in Fig.~2 of the main text can be attributed to the excitation of acoustic strain waves in the thin layer \cite{Zhang2020}. Inspection of the oscillations of the Co data reveals a frequency of $\approx$150\,GHz, corresponding to a time period of $6.\overline6$\,ps. With the speed of sound in Co of $\approx$ 4.7\,nm/ps this corresponds to a distance of $\approx$30\,nm, i.e. twice the thickness of the sample. This suggests that the oscillations are most likely due to an acoustic strain wave launched upon laser excitation that couples to the magnetization. By changing the time range for the fitting algorithm, we confirmed that the oscillations do not influence the extracted values of the demagnetization amplitude, $A$.

\section{Demagnetization time constants}
        
\begin{figure}
    \centering
    \includegraphics[width=0.8\linewidth]{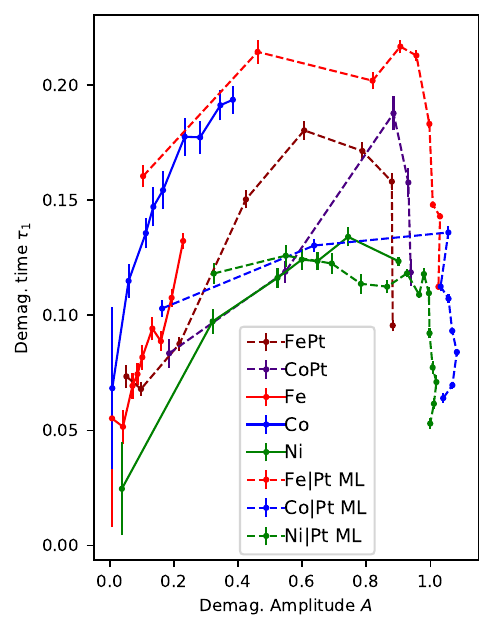}
    \caption{Demagnetization time constant $\tau_1$ for the FM and their respective Pt MLs as a function o the corresponding demagnetization amplitude $A$, where $A=1$ corresponds to full demagnetization.}
	\label{fig:time_vs_A.png}
\end{figure}

Finally, we also want to discuss another important parameter of the demagnetization dynamics: the material-characteristic exponential demagnetization time constant $\tau_1$. As displayed in Fig.~\ref{fig:time_vs_A.png}, we observe an increase of the demagnetization time for small demagnetization amplitudes, $A$, for the pure ferromagnetic films Ni, Fe and Co. Ni exhibits the fastest demagnetization rate ranging between $\tau_1 \approx$25\,fs and 130\,fs, followed by Fe with values between $\tau_1 \approx 50$\,fs and 125\,fs and Co with $\tau_1 \approx$75\,fs and 175\,fs. This is in agreement with Radu et. al \cite{radu_ultrafast_2015}, where the authors show that the element-specific demagnetization time constants depends on the amplitude of the ground state magnetic moments, $\mu$. Differently, the microscopic three temperature model predicts a dominating dependence on the Curie temperature \cite{koopmans_explaining_2010}.  Solution of the coupled differential equations with tabulated material parameters therefore yields a stronger demagnetization rate for Fe compared to Co, which we do not observe. 

As the OISTR process takes place during the presence of the optical laser pulse, one would expect a smaller demagnetization time constants, $\tau_1$, of the multi-component systems compared to the pure elements, again more pronounced for Fe vs. Fe$\vert$Pt/FePt and Co vs. Co$\vert$Pt/CoPt than for Ni vs. Ni$\vert$Pt. Such a decrease in $\tau_1$ is indeed observed for the Co$\vert$Pt multilayer and CoPt alloy versus the Co film, which is also consistent with previous studies \cite{kuiper_spin-orbit_2014, willems_optical_2020}.  For Fe vs FePt, we observe the same trend, the multi-component system again demagnetizes faster. For Fe$\vert$Pt the comparison is less clear as we have almost no data corresponding to the same value of the demagnetization amplitude, $A$, which is due to the fact that for this material pair the ratio of $\Gamma$ is the largest. However, from literature it is known that Fe for $A \approx 0.35$ is $\tau = (230\pm 30)$~fs \cite{Eschenlohr2012, radu_ultrafast_2015}, i.e. slower than what we observe for the Fe$\vert$Pt system. For Ni and Ni$\vert$Pt the time constants are comparable, indicative of the smaller role of OISTR. 
 
\section{Theory}

By studying the dynamics of magnetic moment and charge in the main text we show that the combination of OISTR and spin-orbit coupling (SOC) act together to enhance the demagnetization of Co in CoPt as compared to bulk Co. Compare Fig.~5 of the main text, which displays the results for the case of laser excited CoPt obtained using TDDFT. Here we provide more details of the simulation method and the question of the role of SOC in the OISTR effect. In our simulations the following Hamiltonian is time-propagated\cite{krieger2015,dewhurst2016} 

\begin{eqnarray}
i\frac{\partial \psi_j({\bf r},t)}{\partial t}&=&
\Bigg[
\frac{1}{2}\left(-i{\nabla} +\frac{1}{c}{\bf A}_{\rm ext}(t)\right)^2 +v_{s}({\bf r},t)+ \nonumber \\
&&\frac{1}{2c} {\m \sigma}\cdot{\bf B}_{s}({\bf r},t) + \nonumber \\
&&\frac{1}{4c^2} {\m \sigma}\cdot ({\nabla}v_{s}({\bf r},t) \times -i{\nabla})\Bigg]
\psi_j({\bf r},t)
\label{KS}
\end{eqnarray}
where ${\bf A}_{\rm ext}(t)$ is a vector potential representing the applied laser field, ${\m \sigma}$ the vector of Pauli matrices $(\sigma_x,\sigma_y,\sigma_z)$, and $v_{s}({\bf r},t) = v_{\rm ext}({\bf r},t)+v_{\rm H}({\bf r},t)+v_{\rm xc}({\bf r},t)$ the Kohn-Sham (KS) effective potential. This consists of the external potential $v_{\rm ext}$, the classical electrostatic Hartree potential $v_{\rm H}$ and the exchange-correlation (XC) potential $v_{\rm xc}$, for which we have used the adiabatic local density approximation. Similarly the KS magnetic field is given by ${\bf B}_{s}({\bf r},t)={\bf B}_{\rm ext}(t)+{\bf B}_{\rm xc}({\bf r},t)$ in which ${\bf B}_{\rm ext}(t)$ represents the external magnetic field and ${\bf B}_{\rm xc}({\bf r},t)$ the exchange-correlation~(XC) magnetic field. The final term of Eq.~\ref{KS} is the spin-orbit coupling term. The Kohn-Sham orbitals, $\psi_j$, are two component Pauli spinors. From these Kohn-Sham states the magnetization density can be calculated as

\begin{eqnarray}\label{st}
{\bf m}({\bf r},t)=\sum_j \psi^*_j({\bf r},t){\m \sigma}\psi_j({\bf r},t),
\end{eqnarray}

In the absence of SOC, i.e. if the final term of Eq.~\ref{KS} is set to zero, the wave-functions are pure spinors and the off-diagonal elements of the density matrix strictly zero. This also implies that the total spin is a good quantum number. The results obtained by setting the SOC term to zero in Eq.~\ref{KS} are shown in Fig.~5 (main text), panels a) and b). It is clear from this figure that the local moment lost on the Co atom corresponds to the moment gained on the Pt atom, leading to a zero loss in total moment (some moment is also lost to the highly delocalized high energy states). The reason for this, see Fig.~5~b), is the transfer of Pt minority electrons to empty Co minority states; the loss in Pt nearly exactly equals the gain in Co minority charge. Similarly, Co majority electrons are excited to empty Pt majority states. This is the phenomena of OISTR and in the absence of SOC and the sole mechanism that can cause a change in the local moments of Co and Pt.

Inclusion of SOC allows the spins channels to mix (i.e. off diagonal density matrix in spin space are non-zero), and the results obtained are shown in Fig.~5, panels c) and d). The magnetic moment on Co and Pt atoms now both reduces due to the spin-orbit induced transfer of charge from majority to minority states, however the two competing processes of OISTR and SOC leads to a delayed response of Pt compared to Co, similar to that which has been observed in resonant spectroscopy experiments in a FeNi alloy \cite{Mathias2012,Jana2017}. Comparison of the charge dynamics shown in Fig.~5~d) and b) demonstrates that now more Pt minority charge is transferred to Co minority states. Thus SOC further aids OISTR.

\bibliography{z_bib.bib}